\documentclass[12pt]{article}
\usepackage{amsmath}

\usepackage{bbm}
\usepackage{xcolor}
\usepackage{graphicx}
\usepackage{url}
\usepackage{lscape,verbatim}
\usepackage{xcolor} 
\usepackage{graphics,amsmath,pstricks}
\usepackage{amssymb,enumerate}
\usepackage{amsbsy,amsmath,amsthm,amsfonts, amssymb}
\usepackage{graphicx, rotate, array}
\graphicspath{ {./images/} }
\usepackage{multirow}
\usepackage{float}
\usepackage{natbib}
\usepackage{mathrsfs}
\usepackage{algorithm} 
\usepackage{algpseudocode} 
\usepackage[english]{isodate} 
\usepackage{multirow}
\cleanlookdateon
\theoremstyle{plain}
\newtheorem{lemma}{Lemma}
\newtheorem{theorem}{Theorem}
\usepackage{listings}
\usepackage{subcaption} 
\usepackage{booktabs} 
\usepackage{ifthen}
\usepackage{pstricks}
\usepackage[normalem]{ulem}
\usepackage{titlesec}

\newtheorem{assumption}{Assumption}

\newboolean{DEBUG}
\setboolean{DEBUG}{true}

\newrgbcolor{amycolor}{.5 .1 .99}
\ifthenelse {\boolean{DEBUG}}
{\newcommand{\amy}[1]{{\amycolor{[\@Amy: #1]}}}}
{\newcommand{\amy}[1]{}}

\newrgbcolor{davidcolor}{.1 .7 .1}
\ifthenelse {\boolean{DEBUG}}
{\newcommand{\david}[1]{{\davidcolor{[\@David: #1]}}}}
{\newcommand{\david}[1]{}}

\newrgbcolor{sarahcolor}{.7, .1, .1}
\ifthenelse {\boolean{DEBUG}}
{\newcommand{\sarah}[1]{{\sarahcolor{[\@Sarah: #1]}}}}
{\newcommand{\sarah}[1]{}}

\begin{document}

\def\spacingset#1{\renewcommand{\baselinestretch}%
{#1}\small\normalsize} \spacingset{1.1}

\title{\bf Estimating Fold Changes from Partially Observed Outcomes with Applications in Microbial Metagenomics}
  \author{David S Clausen, Sarah Teichman, and
    Amy D Willis\thanks{
      The authors gratefully acknowledge support from the National Institute of General Medical Sciences (R35 GM133420). Correspondence: adwillis@uw.edu} \\
    Department of Biostatistics, University of Washington}
  \maketitle

\begin{abstract}
We consider the problem of estimating fold-changes in the expected value of a multivariate outcome observed with unknown sample-specific and category-specific perturbations. This challenge arises in high-throughput sequencing studies of the abundance of microbial taxa because microbes are systematically over- and under-detected relative to their true abundances. Our model admits a partially identifiable estimand, and we establish full identifiability by imposing interpretable parameter constraints. To reduce bias and guarantee the existence of estimators in the presence of sparse observations, we apply an asymptotically negligible and constraint-invariant penalty to our estimating function. We develop a fast coordinate descent algorithm for estimation, and an augmented Lagrangian algorithm for estimation under null hypotheses. We construct a model-robust score test and demonstrate valid inference even for small sample sizes and violated distributional assumptions. The flexibility of the approach and comparisons to related methods are illustrated through a meta-analysis of microbial associations with colorectal cancer. 
\end{abstract}

\section{Introduction}

Communities of microorganisms (microbiomes) are studied in many scientific disciplines, including in multiple areas of human health and disease \citep{young2017role,yamashita2017oral,shreiner2015gut,moffatt2017lung}. Modern microbiome research leans heavily on high-throughput sequencing techniques,
which enable researchers to
identify large numbers of microbial taxa on the basis of genetic material detected in a given environment.
In this paper, we consider the problem of making statistical statements about changes in the ``absolute'' concentrations of microbes in an environment using only the ``relative'' information provided by marker gene and whole genome shotgun sequencing. Our approach also accounts for the observation that different microbes are consistently over- or under-counted \citep{McLaren:2019cn}.

\begin{figure}
\begin{center}
\centerline{\includegraphics[width=0.75\textwidth]{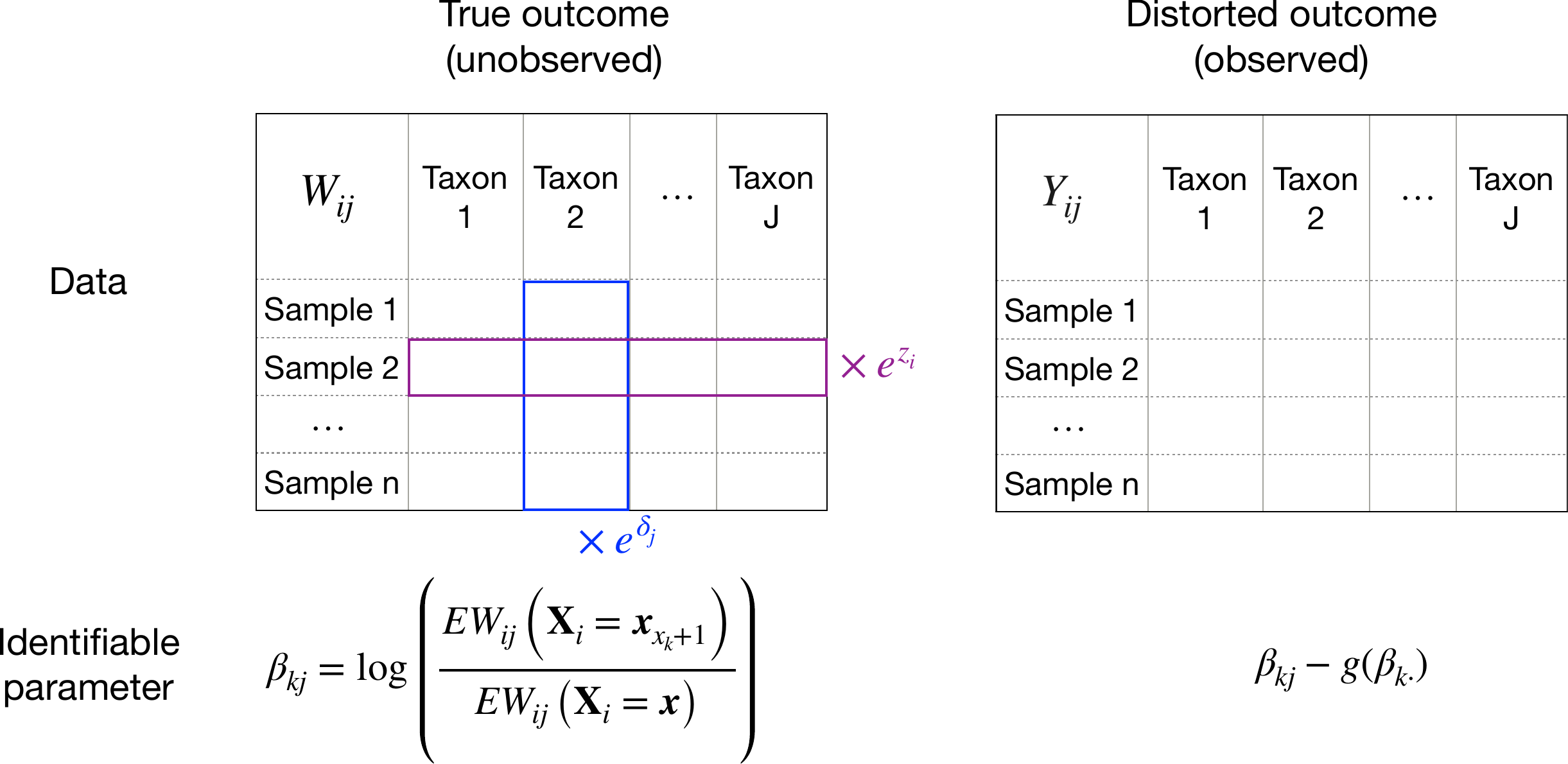}}
\caption{Motivated by differential abundance in microbiome studies, we consider the general problem of estimating fold-differences in the mean of a multivariate outcome in the presence of measurement error. Specifically, when the expected value of the true outcome $W_{ij}$ is multiplicatively distorted by unknown sample-specific ($e^{z_i}$) and category-specific ($e^{\delta_j}$) terms, we show that $e^{\beta_{kj}} = \mathbb{E}W_{ij}(\mathbf{x}_{x_k + 1})/\mathbb{E}W_{ij}(\mathbf{x})$ is not identifiable. However, we show that $\beta_{kj} - g(\beta_{k \cdot})$ is identifiable for appropriate choices of $g: \mathbb{R}^J \rightarrow \mathbb{R}$, a centering function. Reasonable choices include mean-centering, reference group-centering, and smoothed median-centering. The choice of centering function impacts the interpretation of the resulting estimates, but our proposed estimation procedure is $g$-invariant. }  \label{fig:overview}
\end{center}
\end{figure}

We introduce a method for estimation and inference on ratios of
means of a
nonnegative multivariate outcome observed only
up to unknown sample-specific and category-specific perturbations (Figure \ref{fig:overview}). 
To our knowledge, this problem has not been studied in the statistical literature. 
Unknown perturbations are addressed by microbiome data analysts via a range of regression models that typically rely on log-based transformations. These include ANCOM-BC2, 
which delineates zero-valued outcomes into ``structural'' and non-structural zeroes prior to transformation and modeling \citep{lin2024multigroup};
and ALDEx2, which employs a Bayesian approach to impute
nonzero values
before transformation \citep{fernandes2014unifying}.
ANCOM-BC2 and ALDEx2 fall into the larger category of ``differential abundance''
methods,
which include DESeq2, edgeR, metagenomeSeq, IFAA,
and corncob, as well as many others (see \citet{nearing2022microbiome} for an empirical comparison). DESeq2 and edgeR were originally developed for RNAseq data and 
address excess-Poisson dispersion via a negative binomial likelihood \citep{love2014moderated, robinson2010edger}.
By contrast, metagenomeSeq was developed specifically for microbiome analysis and
employs a zero-inflated Gaussian model motivated by the high degree of sparsity in microbiome datasets \citep{paulson2013metagenomeseq}.
IFAA fits a log-linear zero-inflated Gaussian model by identifying a reference taxon whose abundance is conditionally independent of covariates \citep{li2021ifaa}, and  
corncob \citep{martin2020modeling} models both taxon relative abundances and dispersion of
relative abundances via a beta-binomial model. Although the regression models above
are all considered to be differential abundance methods \citep{nearing2022microbiome}, they do not target a common estimand and hence do not produce strictly comparable inference. 

Our manuscript is structured as follows: we introduce our model and establish identifiability of equivalence classes of parameters in Section \ref{sec:model}. We discuss estimation via a generalized estimating equations framework in Section \ref{sec:estimation}, desirable asymptotic properties of our estimator in Section \ref{sec:theory}, and inference via model misspecification-robust testing in Section \ref{sec:inference}. We present simulation results pertaining to Type 1 and 2 error rates in Section \ref{sec:simulations} and consider a data analysis in Section \ref{sec:data}. We conclude with a discussion in Section \ref{sec:discussion}. Open-source software implementing the method is available at \url{github.com/statdivlab/radEmu}, and additional details and proofs are available in Supporting Information (SI). 

\section{Model} \label{sec:model}

\subsection{Motivation}

Our work is motivated by settings in which an ``ideal'' dataset $W_{ij} \geq 0$ (samples $i = 1, \dots, n$ and categories $j = 1, \dots, J$) is not observed, but a distorted proxy $Y_{ij} \geq 0$ is available.
Observing $Y$ rather than $W$ represents a loss of information in two critical ways (Figure \ref{fig:overview}). Firstly, the magnitude of $Y_{i} \in \mathbb{R}^{J}_{\geq 0}$ is uninformative with respect to the magnitude of $W_i \in \mathbb{R}^{J}_{\geq 0}$. In addition, the composition of $Y_{i}$ may be systematically distorted relative to $W_{i}$ as a result of differential detection. That is, observations from category $j_1$ may be more readily observed than observations from category $j_2$, and the magnitude of the detection effects are unknown. Our goal is study the estimability of fold-differences in
$\mathbb{E}W_{\cdot j}$ across covariates, that is, $\mathbb{E}W_{\{i: X_i = x\} j}/\mathbb{E}W_{\{i: X_i = x'\} j}$. 

In our motivating example of microbial communities, $W_{ij}$ represents the (unobserved) cell or DNA concentration of microbial taxon $j$ in sample $i$, which are prohibitively time- and labor-intensive to observe for all taxa $j=1, \ldots, J$ \citep{williamson2022multiview}.
However, it is relatively straightforward to perform high-throughput sequencing to produce data $Y_{ij}$, such as the (observed) read coverage of taxon $j$ in sample $i$ in shotgun sequencing, or the number of times sequence variant $j$ is observed in sample $i$ based on amplicon sequencing. 
In the microbiome meta-analysis considered in Section \ref{sec:data}, we are interested in estimating the fold-difference in the mean cell concentration of taxon $j$ when comparing cancer-diagnosed cases to otherwise similar non-cancer controls, using shotgun sequencing data.

\subsection{Model statement and parameter identifiability} \label{model_statement} 

Suppose that 
\begin{align} \label{eq:true_matrix}
	\log\mathbb{E}[\mathbf{W}|\mathbf{X}] = \mathbf{X}\beta. 
\end{align}
for $\mathbf{W} \in \mathbb{R}^{n \times J}_{\geq 0}$ and $\mathbf{X} \in \mathbb{R}^{n \times p}$. Our target of inference is $\beta \in \mathbbm{R}^{p\times J}$. Unfortunately, however, we do not observe $\mathbf{W}$, and so we cannot directly estimate $\beta$ via \eqref{eq:true_matrix}. Instead, we observe a perturbed outcome $\mathbf{Y} \in \mathbb{R}^{n \times J}_{\geq 0}$ such that
\begin{align} \label{eq:obs_matrix}
	\log\mathbb{E}[\mathbf{Y}|\mathbf{X}] = \mathbf{z}\mathbf{1}_J^T + \mathbf{1}_n \delta^T + \mathbf{X}\beta 
\end{align}
where sample-specific scalings $\mathbf{z} \in \mathbb{R}^{n \times 1}$ and category-specific detection effects $\delta \in \mathbb{R}^{J \times 1}$ are unknown (see Figure \ref{fig:overview}). Hence, the mean of $Y_{ij}$ is equal to the mean of $W_{ij}$ subject to scaling by the sample-specific factor $\exp(z_i)$ and category-specific factor $\exp(\delta_j)$.

As our target of inference is $\beta,$ we consider $\mathbf{z}$ and $\delta$ as nuisance parameters. $\delta$ is straightforward to address if $\mathbf{1}_n$ lies in the column space of $\mathbf{X}$, as is typically the case in practice. Without loss of generality, we consider $\mathbf{X}$ to have the form $[\mathbf{1}_n ~\mathbf{D}]$, and rewrite model (\ref{eq:obs_matrix}) as follows:
\begin{align} \label{eq:betastar}
\log\mathbb{E}[\mathbf{Y}|\mathbf{X}] = \mathbf{z}\mathbf{1}_J^T + \mathbf{X}\beta^{\star}
\end{align}
for $\beta^{\star} = [\beta_0 + \delta, \beta_1, \dots, \beta_{p - 1}]^T$, for $\beta_{k} \in \mathbb{R}^{J \times 1}$ the $k+1$-th row of $\beta$.
In particular, because $\beta_0^{\star}$ depends on the nuisance detection effects $\delta$, it has no scientifically meaningful interpretation. We show below that while $\beta_{kj}$ for $k \geq 1$ is not identifiable from data $\mathbf{Y}$, $\beta_{kj} - g(\beta_{k})$ is identifiable for appropriate centering functions $g$.

We now turn our attention to $\mathbf{z}$, which is not identifiable under our model because for any $\alpha \in \mathbb{R}^p$, we have that
\begin{align} \label{explicit_nonident}
\mathbf{z} \mathbf{1}_J^T + \mathbf{X}(\beta^{\star} +\alpha \mathbf{1}_J^T) & = \mathbf{z}^{\dagger}\mathbf{1}_J^T + \mathbf{X} \beta^{\star}
\end{align}
for $\mathbf{z}^{\dagger} := (\mathbf{z} + \mathbf{X}\alpha)$.
Hence if $\beta^{\dagger} = \beta^{\star} +\alpha \mathbf{1}_J^T$ for some $\alpha \in \mathbb{R}^p$, we can always find a $z_i^{\dagger} \in \mathbb{R}$ such that
\begin{align} \label{non_identifiability}
\log\mathbb{E}[\mathbf{Y}|\mathbf{X}, \beta^{\star}, \mathbf{z}]  = \log \mathbb{E}[\mathbf{Y}| \mathbf{X}, \beta^{\dagger}, \mathbf{z}^{\dagger}],
\end{align}
and therefore mean function \eqref{eq:betastar} is not identifiable in $\beta^{\star}$ and $\mathbf{z}$.
However, mean function \eqref{eq:betastar} is \textit{partially identifiable} because we can define equivalence classes 
such that for $\beta^{\star}$ and $\beta^{\dagger}$ in the same equivalence class, there always exist $\mathbf{z}$ and $\mathbf{z}^{\dagger}$ such that
(\ref{non_identifiability}) holds, and for $\beta^{\star}$ and $\beta^{\dagger}$ in distinct classes, there never exist such $\mathbf{z}$ and $\mathbf{z}^{\dagger}$. We prove the following result in Supporting Information (Section \ref{appA}). 

\begin{lemma} \label{thm:identifiability}
For arbitrary $b \in \mathbb{R}^{p\times J}$, define $$G_{b} = \{\beta^{\star} \in \mathbbm{R}^{p \times J}: \beta^{\star} = b + \alpha\mathbf{1}_J^T \text{ for some } \alpha \in \mathbb{R}^p\}.$$
If $\mathbf{X}$ has full column rank, for any $\beta^{\star}, \beta^{\dagger} \in \mathbb{R}^{p\times J}$ there exist $\mathbf{z} \in \mathbb{R}^n$ and $\mathbf{z}^{\dagger} \in \mathbb{R}^n$ such that
under model (\ref{eq:betastar}),
$\log\mathbb{E}[\mathbf{Y}|\mathbf{X}, \beta^{\star}, \mathbf{z}] = \log\mathbb{E}[\mathbf{Y}|\mathbf{X}, \beta^{\dagger}, \mathbf{z}^{\dagger}]$ if and only if $\beta^{\dagger} \in G_{\beta}$.
\end{lemma} 

Lemma \ref{thm:identifiability} demonstrates that while we cannot uniquely estimate $\beta_1, \ldots, \beta_{p-1}$ from data generated according to (\ref{eq:obs_matrix}),
we can estimate them up to the addition of constant terms.
That is, while we cannot estimate differences across levels of covariate $X_{\cdot k}$ in the log-means of $W_{\cdot j}$ given only data $\mathbf{Y}$, we \textit{can} estimate relative differences in the log-mean of $W_{\cdot j_1}$ compared to $W_{\cdot j_2}$. In genomics studies, this represents a small loss of information because researchers typically aim to identify categories that show markedly different abundances across covariate groups of interest. That is, even if all categories differ in abundance across covariate groups, the categories of \textit{most} interest are those with the largest differences. Therefore, even if we cannot identify $\beta_1, \ldots, \beta_{p-1}$ uniquely, estimating their equivalence classes can answer questions about differences in unobserved true abundances $\mathbf{W}$ across groups defined by $\mathbf{X}$.

We estimate the equivalence class of $\beta_1, \ldots, \beta_{p-1}$ by imposing identifiability constraints of the form $g(\beta_k) = 0$ such that each $G_{\beta}$ contains only a single unique element satisfying these constraints. As is typical of overparameterized models (e.g., one-way ANOVA in the absence of constraints on category means $\mu + \alpha_i$), the choice of constraint changes the interpretation of the parameters. For example, under constraint $\beta_{kj'} = 0$, $\beta_{kj}$
represents $\log \mathbb{E}[W_{\cdot j}|X_{\cdot} = \mathbf{x}_{x_k+1}] - \log \mathbb{E}[W_{\cdot j}|X_{\cdot} = \mathbf{x}]$ minus the corresponding
quantity for taxon $j'$, where $\mathbf{x}_{x_k+1}$ is equal to $\mathbf{x}$ in all elements excepting the $k$th, which is increased by 1 unit. We equivalently interpret $e^{\beta_{kj}}$ as a fold-difference (multiplicative difference). 

\vspace{0.25cm}
\textit{
	\textbf{Example 1: } We contrast two scenarios: assuming that $\beta_{kj'} = 0$, and imposing the constraint that $\beta_{kj'} = 0$. In situations where a ``reference category'' is known to have $\beta_{kj'} = 0$, the constraint $\beta_{kj'} = 0$ means that $\beta_{kj}$ can be interpreted as the log-fold difference in $\mathbb{E}W_{\cdot j}$ across groups that differ by one unit in $X_{\cdot k}$. Therefore, knowledge of a reference category allows interpretation of parameters as fold-differences in mean abundances $W$. 
	In contrast, when no such category is known, $\beta_{kj}$ is not identifiable from data $\mathbf{Y}$. Imposing the constraint $\beta_{kj'} = 0$ means that $\beta_{kj}$ can be interpreted as the log-fold difference in $\mathbb{E}W_{\cdot j}$ across groups that differ by one unit in $X_{\cdot k}$, minus the corresponding log-fold difference in $\mathbb{E}W_{\cdot j'}$. That is, $\beta_{kj}$ can only be interpreted relative to $\beta_{kj'}$.}
	\vspace{0.25cm}

Identifiability can also be imposed relative to all categories, rather than relative to a specific category. 
We first consider constraining $\text{median}_{j=1, \ldots, J}(\beta_{kj}) = 0$ for all $k$. The robustness of the median is attractive, 
as no 
single large effect size 
will impact the median 
(Supplementary Figure 1). However, the median is not a differentiable function of $\beta_k$ (its total derivative is not guaranteed to exist), which
substantially complicates inference (see Section \ref{sec:inference}).
Accordingly, we propose a smooth approximation to the median that
shares its robustness properties: the pseudo-Huber loss  $g_{p}(\beta_k) = \arg \min_c \sum_{j=1}^J h_\gamma\left(\beta_{kj} - c\right)$ for $h_\gamma(x) = \gamma^2 \left(\sqrt{1 + \left( x/\gamma \right)^2} -1 \right)$ for prespecified parameter $\gamma >0$ \citep{barron2019general}. The pseudo-Huber loss is approximately quadratic in a neighborhood of 0 and asymptotically approaches the function $|x|/\gamma$ for large $x$. The parameter $\gamma$ controls the size of the quadratic region around zero: as $\gamma \downarrow 0$, $g_{p}(\beta_k)$ approaches the median, and as $\gamma \rightarrow \infty$, it approaches the mean. We propose $\gamma = 0.1$ as a reasonable default for many applications, though empirically we find that care is warranted when covariates are on very different scales.
The following example illustrates that enforcing identifiability via the smoothed median constraint does not imply any structural assumptions on $\beta$.

\vspace{0.25cm}
\textit{
	\textbf{Example 2: } Similar to Example 1, if $g_p(\beta_{k})$ is known to equal zero (e.g., $\beta_{k}$ is known to be sparse or symmetric), then $\beta_{kj}$ can be interpreted as the log-fold difference in $\mathbb{E}W_{\cdot j}$ across groups that differ by one unit in $X_{\cdot k}$. Without this assumption, applying the constraint $g_p(\beta_{k})=0$ means that $\beta_{kj}$ should be interpreted as the log-fold difference in $\mathbb{E}W_{\cdot j}$ across groups that differ by one unit in $X_{\cdot k}$ relative to the typical such log-fold difference across taxa.}
\vspace{0.25cm}

Further illustration of the advantages of estimating fold-differences relative to smoothed medians are given in Supplementary Figure 1. 
We propose $g_p$ as the default constraint function and give a detailed example of interpreting partially identifiable estimands without structural assumptions on $\beta$ in Section \ref{sec:data}. That said, any smooth $g: \mathbb{R}^{J}  \rightarrow \mathbb{R}$ satisfying $g(\beta_k + \alpha) =  g(\beta_k) + \alpha$ is possible, and the most appropriate constraint will depend on the scientific setting. 

\section{Estimation} \label{sec:estimation}

\subsection{Estimation under $H_A$} \label{subsec:unconstrained}
We estimate $\beta$ (our target of inference; subject to identifiability constraints) and $\mathbf{z}$ (a nuisance parameter) using an estimating equations approach motivated by a Poisson likelihood with mean function \eqref{eq:betastar}.
To address issues of infinite or undefined MLEs resulting from data separation, which commonly occurs in high-dimensional microbiome sequencing data, we introduce a Firth penalty on $\beta$ derived from a formal equivalence between our model and the multinomial logistic model. 
We develop a fast algorithm to find the maximizer of the penalized likelihood by efficiently solving a sequence of score equations that converge to the score of the penalized likelihood within a neighbourhood of the global maximizer. 
Theorem \ref{thm:normality_g} illustrates some advantages of this proposal: the resultant estimator is normally distributed and consistent for the identifiable parameter $\beta_k - g(\beta_k)$ for any distribution of $Y_{ij}$ satisfying \eqref{eq:betastar}, with no requirement for $Y_{ij}$ to be Poisson distributed. In addition, Sections \ref{sec:inference} and \ref{sec:simulations} demonstrate that the estimator and accompanying hypothesis testing show excellent performance even when data has excess-Poisson variance, and even in small samples.
All algorithms are provided as pseudocode in Supporting Information (Section \ref{weighted_supp}). 

To motivate our estimator, we first introduce a Poisson log likelihood for our model without yet considering the identifiability constraints $g(\beta_k) = 0$.
Letting $\beta^j \in \mathbb{R}^{p \times 1}$ denote the $j$-th column of $\beta$ and $X_i \in \mathbb{R}^{p  \times 1}$ denote the $i$-th row of $\mathbf{X}$, the Poisson log likelihood is
\begin{align}
l_n^{\text{Poisson}} (\beta,z) &= \sum_{j = 1}^J \left[ \sum_{i = 1}^n \Big(Y_{ij} (X_i^T\beta^j + z_i) - \exp (X_i^T\beta^j + z_i)\Big) \right]. \label{likelihood}
\end{align}
For
fixed $\beta$, the maximizing value of $\mathbf{z}$ is available in closed form
\begin{align}
\label{eq:profile_z}
\hat{z}_i & = \log\sum_{j = 1}^J Y_{ij} - \log\sum_{j = 1}^J  \exp (X_i^T\beta^j),
\end{align}
and therefore we can
profile out $\mathbf{z}$ in likelihood (\ref{likelihood}) to give
\begin{align}
l_n^{\text{profile}} (\beta) &= \sum_{j = 1}^J \left[ \sum_{i = 1}^n \Big(Y_{ij} \log  \frac{\exp (X_i^T\beta^j)}{\sum_{j' = 1}^J \exp(X_i^T\beta^{j'})} - \big|\big|Y_{i}\big|\big|_1 \frac{\exp (X_i^T\beta^j)}{\sum_{j' = 1}^J \exp(X_i^T\beta^{j'})}\Big) \right] + C \label{prof_ll}
\end{align}
where $C$ is constant (with respect to $\beta$) 
and $\big|\big|Y_{i}\big|\big|_1 := \sum_{j = 1}^J Y_{ij}$. Note that the profile likelihood (\ref{prof_ll}) is equal (up to a constant) to a multinomial log likelihood with a logistic link.
This accords with both known results about the marginal Poisson distribution of multinomial random variables \citep{birch1963maximum}, and our observations on identifiability of $\beta$ (only $p\times(J - 1)$ parameters can be identified in a multinomial logistic regression of a $J$-dimensional outcome on $p$ regressors). 

Unfortunately, the maximizer of $l_n^{\text{profile}}$ will not be finite if there exists a perfect separating hyperplane \citep{albert1984existence}. To both guarantee the existence of an estimate and reduce its bias, we derive the Firth penalty associated with likelihood (\ref{prof_ll}), observing the impact of the nuisance parameters $\mathbf{z}$ on the correct bias-reducing penalty \citep{kosmidis2011multinomial}.
Fix $j^{\dagger} \in \{1, \ldots, J\}$ and let $\tilde{\beta} := [{\beta^1}^T, \ldots, {\beta^{j^{\dagger} - 1}}^T,{\beta^{j^{\dagger} + 1}}^T, \ldots, {\beta^{J}}^T]^T \in \mathbb{R}^{p(J-1) \times 1}$, that is, a column vector containing stacked columns of $\beta$ but with $\beta^{j^{\dagger}}$ omitted. Let $\tilde{\mathbf{X}}_i = I_J^{[-j^{\dagger}]}  \otimes X_i^T$, where $I_J^{[-j^{\dagger}]}$ indicates a  $J\times J$ identity matrix whose $j^{\dagger}$-th column has been removed and $\otimes$ is the Kronecker product, so $\tilde{\mathbf{X}}_i \in \mathbbm{R}^{J\times p(J-1)}$ and $\tilde{\mathbf{X}} := [\tilde{\mathbf{X}}_1^T,\dots,\tilde{\mathbf{X}}_n^T]^T \in \mathbb{R}^{nJ\times p(J - 1)}$. Define the fitted values $p_i = \exp(\tilde{\mathbf{X}}_i\tilde{\beta})\big / \mathbf{1}_J^T\exp(\tilde{\mathbf{X}}_i\tilde{\beta}) \in \mathbb{R}^J$. Then, the information matrix associated with (\ref{prof_ll}) is
$\tilde{\mathbf{X}}^T\mathbf{V}\tilde{\mathbf{X}}$ where $\mathbf{V}$ is a block diagonal matrix with $i$-th $J \times J$ block equal to $\big|\big|Y_{i}\big|\big|_1(\text{diag}(\mathbf{p}_i) - \mathbf{p}_i\mathbf{p}_i^T)$ with $\text{diag}(\mathbf{p}_i)$ indicating a diagonal matrix with diagonal elements equal to $p_{i1}, \dots, p_{iJ}$ \citep{bull2002modified}. Finally,  we can express the Firth penalized multinomial log likelihood as follows:
\begin{align}
l_n^{\text{penalized}} = l_n^{\text{profile}} + \frac{1}{2}\log \big|\tilde{\mathbf{X}}^T\mathbf{V}\tilde{\mathbf{X}}\big|,  \label{pen_ll}
\end{align}
where the penalty term depends on $\beta$ through $\mathbf{V}$.

Remarkably, the penalized likelihood (\ref{pen_ll}) is invariant to the choice of $j^{\dagger}$.
This is because
the Firth penalty is invariant under smooth reparametrizations for full exponential family models such as the multinomial distribution \citep{jeffreys1946invariant,firth1993bias}, and the above parameterization implies the constraint $g(\beta_k) = \beta_{kj^{\dagger}}=0$. Thus, we can estimate $\beta$ under the ``convenience constraint'' $\tilde{\beta}_{kj^{\dagger}} = 0$ and impose the desired identifiability constraint $g(\beta_k) = 0$ post-hoc by substracting $g(\tilde{\beta}_k)$ from $\tilde{\beta}_k$.
In practice we find that choosing $j^{\dagger}$ to maximize $\sum_{i} \mathbf{1}_{[Y_{ij^{\dagger}}>0]}$ improves the speed of optimization.

To find $\arg \max_{\tilde{\beta}} l_n^{\text{penalized}}$, we apply the iterative maximum likelihood algorithm of \cite{kosmidis2011multinomial}, which maximizes (\ref{pen_ll}) by iteratively solving multinomial score equations in augmented data $\mathbf{Y}^{(t)}$. Specifically, $Y^{(t)}_{ij} = Y_{ij} + h_{ij}/2$ where $h_{ij}$ is the $(i-1)\times J + j$-th diagonal element of the Poisson ``hat'' matrix $\mathbf{V}_p^{1/2}\tilde{\mathbf{G}} (\tilde{\mathbf{G}}^T\mathbf{V}_p\tilde{\mathbf{G}})^{-1}\tilde{\mathbf{G}}^T\mathbf{V}_p^{1/2}$ evaluated at current values of $\beta^{(t)}$ 
for $\mathbf{V}_p$ an $nJ \times nJ$ diagonal matrix with $(i - 1)\times J + j$-th diagonal element equal to $\exp(X_i^T\beta^j + z_i)$ and $\tilde{\mathbf{G}} = \left[ \tilde{\mathbf{X}} : \mathbf{I}_n \otimes \mathbf{1}_J\right] \in \mathbb{R}^{nJ\times (p(J - 1) + n)}$ (i.e., the expanded design matrix for $\beta$, $\tilde{\mathbf{X}}$, side by side with a design matrix for $\mathbf{z}$, $ \mathbf{I}_n \otimes \mathbf{1}_J$).
We compute data augmentations under the restriction on $z$ that $\sum_j \exp(X_i^T\beta^j + z_i) = \big|\big|Y_{i}\big|\big|_1$ by applying (\ref{eq:profile_z}). 
Each iteration $t$ of this algorithm is separable in $z, \beta^1, \ldots, \beta^J$, and we therefore solve each iteration using coordinate descent, applying (\ref{eq:profile_z}) to update $z$ and updating each $\beta^j$ with Fisher scoring followed by a line search.
When a desired convergence threshold is reached, we impose the desired identifiability constraint by centering by $g(\tilde{\beta}_k^{(t_{\text{max}})})$ as described previously to obtain maximizing estimates satisfying $g({\beta}_k) = 0$.

\subsection{Estimation under $H_0$}

In Section \ref{sec:inference}, we present a robust score test for $\beta_{k^{\star}j^{\star}} = 0$ under constraint $g(\beta_{k^{\star}}) = 0$, which requires fitting a model under the restriction $\beta_{k^{\star}j^{\star}} = g(\beta_{k^{\star}})$. We therefore develop an augmented Lagrangian algorithm \citep[p. 514]{jorge2006numerical} to solve this constrained optimization problem. The augmented Lagrangian function is
\begin{align} \label{eqn:auglag}
\mathcal{L}(\beta, z; \rho, u) &:= -l_n^{\text{Poisson}}(\beta, z; \mathbf{Y}^{(t_{\text{max}})}) + u\left[g\left(\beta_{k^{\star}}\right) - \beta_{k^{\star}j^{\star}}\right] + \frac{\rho}{2}\left[g\left(\beta_{k^{\star}}\right)- \beta_{k^{\star}j^{\star}}\right]^2
\end{align}
for $\mathbf{Y}^{(t_{\text{max}})}$ from our model fit under $H_A$.
For sequences $\rho^{(t)}$ and $u^{(t)}$ described below,
we iteratively find maximizers $(\hat{\beta}_{aug}^{(t)},\hat{z}_{aug}^{(t)})$ of (\ref{eqn:auglag}) until $\hat{\beta}_{aug}^{(t)}$
satisfies the constraint $\beta_{k^{\star}j^{\star}} = g(\beta_{k^{\star}})$ to within a prespecified tolerance. 
While ideally (\ref{eqn:auglag}) would be a function of $l^{\text{penalized}}(\beta; \mathbf{Y})$ instead of $l^{\text{Poisson}}(\beta, z; \mathbf{Y}^{(t_{\text{max}})})$,
updating data augmentations within Lagrangian steps adds an additional inner loop,
and furthermore $l^{\text{Poisson}}(\beta, z; \mathbf{Y}^{(t_{\text{max}})})$ is both a good approximation to $l^{\text{penalized}}(\beta; \mathbf{Y})$ and asymptotically equivalent under $H_0$.

The presence of $g(\beta_{k^{\star}})$ in (\ref{eqn:auglag}) complicates optimization, as this term may depend on an entire row of $\beta$ (e.g., if $g = g_p$).
As a result, updating $\beta$ column-by-column (as when fitting models under $H_A$) often results in slow convergence under $H_0$. 
Hence, we instead minimize (\ref{eqn:auglag}) via coordinate descent alternating between \textit{block} updates to $\beta$ via an approximate Newton step and 
 updates to $z$ via equation (\ref{eq:profile_z}). 
  
  For fixed $z$, the augmented Lagrangian (\ref{eqn:auglag}) has second derivative in $\tilde{\beta}$ (the vectorization of $\beta$ with convenience constraint $\beta^{j^{\dagger}} = \mathbf{0}$) given by
 \begin{align}
 \nabla^2 \mathcal{L} &= -\nabla^2 l^{\text{Poisson}} + [u + \rho (g - \beta_{k^{\star}j^{\star}})] \nabla^2g  + \rho (\nabla g - \vec{e}_{[k^{\star}j^{\star}]})(\nabla g - \vec{e}_{[k^{\star}j^{\star}]})^T
 \end{align}
where 
$\vec{e}_{[k^{\star}j^{\star}]}$ is a vector equal to 1 in the position in $\tilde{\beta}$ corresponding to $\beta_{k^{\star}j^{\star}}$ and zero elsewhere. $\nabla^2 l^{\text{Poisson}} $ is block diagonal and hence inexpensive to invert, and $\rho (\nabla g - \vec{e}_{[k^{\star}j^{\star}]})(\nabla g - \vec{e}_{[k^{\star}j^{\star}]})^T$ is
a matrix of rank one. We therefore approximate the Hessian of (\ref{eqn:auglag}) as $-\nabla^2 l  + \rho (\nabla g - \vec{e}_{[k^{\star}j^{\star}]})(\nabla^T g - \vec{e}_{[k^{\star}j^{\star}]})$, and compute its inverse as a rank-one update to
$-(\nabla^2 l)^{-1}$ via the
Sherman-Morrison identity \citep{sherman1950adjustment}. For additional details, see Supporting Information (Section \ref{weighted_supp}). 

In each iteration of the outer loop of our algorithm, we update $\rho$ and $u$ and obtain an approximate solution to 
(\ref{eqn:auglag}). The update to $u$ is given by $u^{(t + 1)} = \rho^{(t)}[g(\beta_{k^{\star}}^{(t)})- \beta^{(t)}_{k^{\star}j^{\star}}]$ We update $\rho$ if we have not progressed sufficiently toward a feasible solution in the 
past step: if $|g(\beta_{k^{\star}}^{(t)})- \beta^{(t)}_{k^{\star}j^{\star}}| > \kappa|g(\beta_{k^{\star}}^{(t-1)})- \beta^{(t-1)}_{k^{\star}j^{\star}}|$, we set $\rho^{(t + 1)} = \tau\rho^{(t)}$ for 
$\kappa \in (0,1)$ and $\tau >1$; otherwise, we let $\rho^{(t + 1)} = \rho^{(t)}$. We find that default values $\kappa = 0.8$ and $\tau = 2$ yield stable optimization in a broad range of settings. Lower $\kappa$ and higher $\tau$ yield potentially faster optimization but at the cost of stability. In particular, if $\rho$ grows too large, the augmented Lagrangian subproblem may become ill-conditioned.

We evaluate convergence with two criteria. Firstly, we require (primal) feasibility be satisfied to within a chosen tolerance: $|g(\beta_{k^{\star}}) - \beta_{k^{\star}j^{\star}}|< \varepsilon$ for a small $\varepsilon >0$. We additionally require that $\max_{k,j} |\beta^{(t)}_{kj} - \beta^{(t - 1)}_{kj}| < \Delta$ for a small $\Delta>0$. 

\section{Asymptotic properties} \label{sec:theory}

We now present consistency, normality and efficiency results for our proposed method. These results demonstrate that if mean model \eqref{eq:obs_matrix} holds, our estimator with identifiability constraint $g(\hat{\beta}_k) = 0$ converges to the identifiable parameter $\beta - g(\beta_k)$ with a normal limiting law. 
We also give a stronger condition under which our estimator achieves the asymptotic lower bound on variance. 
We require no parametric distributional assumptions on $\mathbf{Y}$ (e.g., Poisson, zero-inflated Negative Binomial, etc.) for any of the below results.  
The absence of parametric assumptions is critical because we believe that parametric models cannot fully capture the characteristics of high-throughput sequencing data. 

We begin by considering the identifiability constraint $g(\beta_k) = \beta_{kJ}$, and then generalize these results to more general smooth constraint functions $g$. 
Under the parametrization $g(\beta_k) = \beta_{kJ}$, the model we obtain from profile likelihood (\ref{prof_ll}) is a multinomial logistic regression with 
reference category $J$. 



We begin by showing that the maximizer of the population profile likelihood is the same $\beta$ given by the mean model \eqref{eq:obs_matrix}, subject to identifiability. Proofs for all results can be found in Supporting Information Section \ref{appA}. 

\begin{assumption} \label{assumption:mean}
	Let $(X, Y, Z) \overset{iid}{\sim} P^{(0)}$ for unknown true data generating mechanism $P^{(0)}$ with bounded support in $\mathbb{R}^p \times \mathbb{R}^J \times \mathbb{R}.$ Then $P^{(0)}$ satisfies $\log \mathbbm{E}[Y|X,Z;\beta^{(0)}] = Z\mathbf{1}_J + {\beta}^{(0)T}X$ for ${\beta}^{(0)} \in \mathbb{R}^{p \times J}$ with $J$-th column equal to $\mathbf{0}_J$. 
\end{assumption}

\begin{assumption}	\label{assumption:pd}
	$P^{(0)}$ satisfies $\Sigma_X := Cov(X)$ is positive definite. 
\end{assumption}

\begin{lemma} \label{lemma:maximizer}
	Under Assumptions \ref{assumption:mean} and \ref{assumption:pd}, $\beta^{(0)}$ is the unique maximizer of the population criterion $M(\beta) := \mathbbm{E}_{P^{(0)}}\hspace{0.05cm}l_i^{\text{profile}}(\beta)$. 
\end{lemma}

We then show that our estimator, which maximizes the Firth-penalized sample likelihood, converges to $\beta^{(0)}$. While we consider a random design, it is straightforward to generalize this result to a fixed design using by placing an eigenvalue condition on the design matrix \citep{chen1999strong}.

\begin{assumption} \label{assumption:bounded}
${\beta}^{(0)}$ lies in a closed, bounded subset $\mathbf{\Omega}$ of $\mathbb{R}^{p\times J}$
\end{assumption}

\begin{theorem} \label{thm:consistency}
Under Assumptions \ref{assumption:mean} -- \ref{assumption:bounded}, the maximum penalized likelihood estimator (MPLE) $\hat{\beta}^{(n)}$ is consistent. That is, $\hat{\beta}^{(n)}  \xrightarrow{\text{p}} \beta^{(0)}$. 
\end{theorem}

The demonstration of asymptotic normality is simplified by considering the  vector-valued parametrization $\tilde{\beta} \in \mathbb{R}^{p(J-1) \times 1}$ of the matrix-valued parameter $\beta \in \mathbb{R}^{p \times (J-1)}$. 
For $X \in \mathbb{R}^{p}$, we move between the vector notation $\tilde{\beta}$ and matrix notation $\beta \in \mathbb{R}^{p \times (J-1)}$ using the equality $\tilde{X}\tilde{\beta} = {\beta}^{T}X$ for $\tilde{X} = I_J^{[-J]} \otimes X^T \in \mathbb{R}^{J \times p(J-1)}$, noting that we drop the $J$-th column of $\beta$ via $g(\beta_k) = \beta_{kJ}$. 

\begin{theorem} \label{thm:normality}
Under Assumptions \ref{assumption:mean} -- \ref{assumption:bounded}, the MPLE $\sqrt{n}\left(\hat{\tilde{\beta}}^{(n)} - \tilde{\beta}^{(0)}\right) \overset{\mathcal{D}}{\rightarrow} \mathcal{N}(\mathbf{0},\Sigma)$ for positive definite $\Sigma$. 
\end{theorem}

Given the above results for $\hat{\beta} \in \mathbb{R}^{p \times J}$ with $\hat{\beta}_{\cdot J} = \mathbf{0}_p$, expanding the above results to generic constraint function $g(\hat{\beta}_{k}) = \mathbf{0}_p$ only requires the smoothness of $g$. Choosing $g$ to satisfy $g(\mathbf{b}+ a) = g(\mathbf{b}) + a$ is not necessary for Theorem \ref{thm:normality_g}. That said, the identifiability result of Lemma \ref{thm:identifiability} as it relates to estimating the parameter associated with the unobserved data $\mathbf{W}$ means this additional requirement on $g$ is a useful practical restriction. 



\begin{theorem} \label{thm:normality_g}
Fix $k \in \{1, \ldots, p\}$ and $j \in \{1, \ldots, J\}$. 
Define $\hat{\gamma}^{(n)} := \hat{\beta}_{kj}^{(n)} - g(\hat{\beta}_k^{(n)})$ for a smooth function $g: \mathbb{R}^{J} \rightarrow \mathbb{R}$.  
Under Assumptions \ref{assumption:mean} -- \ref{assumption:bounded}, $\hat{\gamma}^{(n)} \xrightarrow{\text{p}} \beta^{(0)}_{kj} - g(\beta^{(0)}_{k})$, and $\sqrt{n}\left(\hat{\gamma}^{(n)} - (\beta^{(0)}_{kj} - g(\beta^{(0)}_{k}))\right) \overset{\mathcal{D}}{\rightarrow} N\left(0, H^T\Sigma H\right)$ for some matrix $H$.
\end{theorem}

We conclude this section by showing that our proposed estimator attains the smallest possible asymptotic variance when the variance of the observations is
proportional to either Poisson or Multinomial variance. While we do not necessarily believe either variance assumption, this result demonstrates when our proposed estimator is not just asymptotically correct, but also optimal within a desirable class of estimators. 

\begin{theorem} \label{thm:efficiency}
	Under Assumptions \ref{assumption:mean} -- \ref{assumption:bounded} and either $\text{Var}(Y| X,Z;\beta^{(0)}) = \phi \times$ $\text{diag} \left\{ \mathbbm{E}[Y|X,Z;\beta^{(0)}]\right\}$ or $\text{Var}(Y| X,Z;\beta^{(0)}) = \phi \tilde{\mathbf{V}}$ for unknown constant $\phi$, $\hat{\beta}_n$ is asymptotically efficient among asymptotically linear estimators of $\beta^{(0)}$. 
\end{theorem}

\section{Inference} \label{sec:inference} 

We now discuss evaluating evidence for a difference in
$\mathbb{E} W_{\cdot j}$ across levels of a covariate $X_{\cdot k}$, holding other covariates fixed. We explore the simple but common setting of testing $\overset{\circ}{H}_0: \beta_{kj} = 0$, but the approach can be generalized to test more general hypotheses of the form $\mathbf{A}\beta= \mathbf{C}$ for fixed matrices $\mathbf{A} \in \mathbb{R}^{h \times p}$ and $\mathbf{C} \in \mathbb{R}^{h \times J}$. 
We are not
able to test $\overset{\circ}{H}_0$ directly, as $\beta$ is not fully identified.
However,
we are able to test $H_0^{(g)}: \beta_{kj} = g(\beta_{k})$,
or equivalently
\begin{align}
 H_0: \text{ } \log \mathbb{E}[ W_{\cdot j}|X_{\cdot } &= \mathbf{x}_{x_k+1}] - \log \mathbb{E}[W_{\cdot j}|X_{\cdot} = \mathbf{x}] \notag \\
 &= g\left( \log \mathbb{E}[W_{\cdot j}|X_{\cdot} = \mathbf{x}_{x_k+1}] - \log \mathbb{E}[W_{\cdot j}|X_{\cdot} = \mathbf{x}]\right) \notag 
\end{align}
Thus, the interpretation of the test depends on the identifiability constraint. 
For example, we may test that a log-fold difference equals the smoothed median of such differences via $g(\beta_k) = g_p(\beta_k)$, or that a log-fold difference equals the log-fold difference of category 1 via $g(\beta_k) = \beta_{k1}$. 

To construct a test of $H_0^{(g)}$, we restate $H_0^{(g)}$ as follows: $H_0^{(g)}: h(\beta) = 0$
with $h(\beta) = \beta_{kj} - g(\beta_k)$.
We test nulls of this form via a robust score test, using the test statistic given in \citet{white1982maximum}:
\begin{align} \label{robust_score}
T_{RS} =S_{H_0}^T I_{H_0}^{-1}F_{H_0}^T (F_{H_0}I_{H_0}^{-1}D_{H_0}I_{H_0}^{-1}F_{H_0}^T)^{-1}F_{H_0}I_{H_0}^{-1}S_{F_{H_0}} 
\end{align}
where $S_{H_0}$ is the score evaluated at the maximum likelihood estimate $\hat{\beta}_{H_0}$ under the null, $I_{H_0}$ is a consistent estimate
of the information matrix under the null, 
$F_{H_0}$ is given by $\frac{\partial h}{\partial \beta^T}$ evaluated at $\hat{\beta}_{H_0}$ and $D_{H_0} = \sum_{i = 1}^n s_i(\hat{\beta}_{H_0})s_i(\hat{\beta}_{H_0})^T$
where $s_i(\beta) = \tilde{X}_i^T(Y_i - \big|\big|Y_{i}\big|\big|_1 p_i(\beta))$ is the contribution to the score of the $i$th observation. 
In practice, we use a score statistic $T'_{RS}$ equal to  $\frac{n}{n - 1}T_{RS}$, where the term in front of $T_{RS}$ is an adjustment due to \citet{guo2005small} that 
reduces the small-sample conservatism of the test. 
Under $H_0$, $T'_{RS}$ is asymptotically $\chi^2$-distributed with one degree of freedom; empirically, we observe that this test provides error rate control in large samples and conservative inference in small to moderate samples (Section \ref{sec:simulations}).
We also consider a robust Wald approach to testing hypotheses of the form $H_0^{(g)}: h(\beta) = 0$. The robust Wald statistic is given by 
\begin{align}
T_{RW} = h(\beta)^T(F_{H_A}^T I_{H_A} D_{H_A}^{-1} I_{H_A} F_{H_A})^{-1}h(\beta)
\end{align}
\noindent where $F_{H_A}$, $I_{H_A}$, and $D_{H_A}$ are as above but evaluated under the full model. Since robust Wald testing does not require 
refitting a model under a null hypothesis, it offers a computational advantage over robust score testing. However, as explored in Section \ref{sec:simulations}, 
robust Wald tests fail to adequately control Type 1 error in small samples. In contrast, robust score tests controlled error rates in all investigated scenarios.

\begin{figure} 
	\begin{center}
	\centerline{\includegraphics[width=\textwidth]{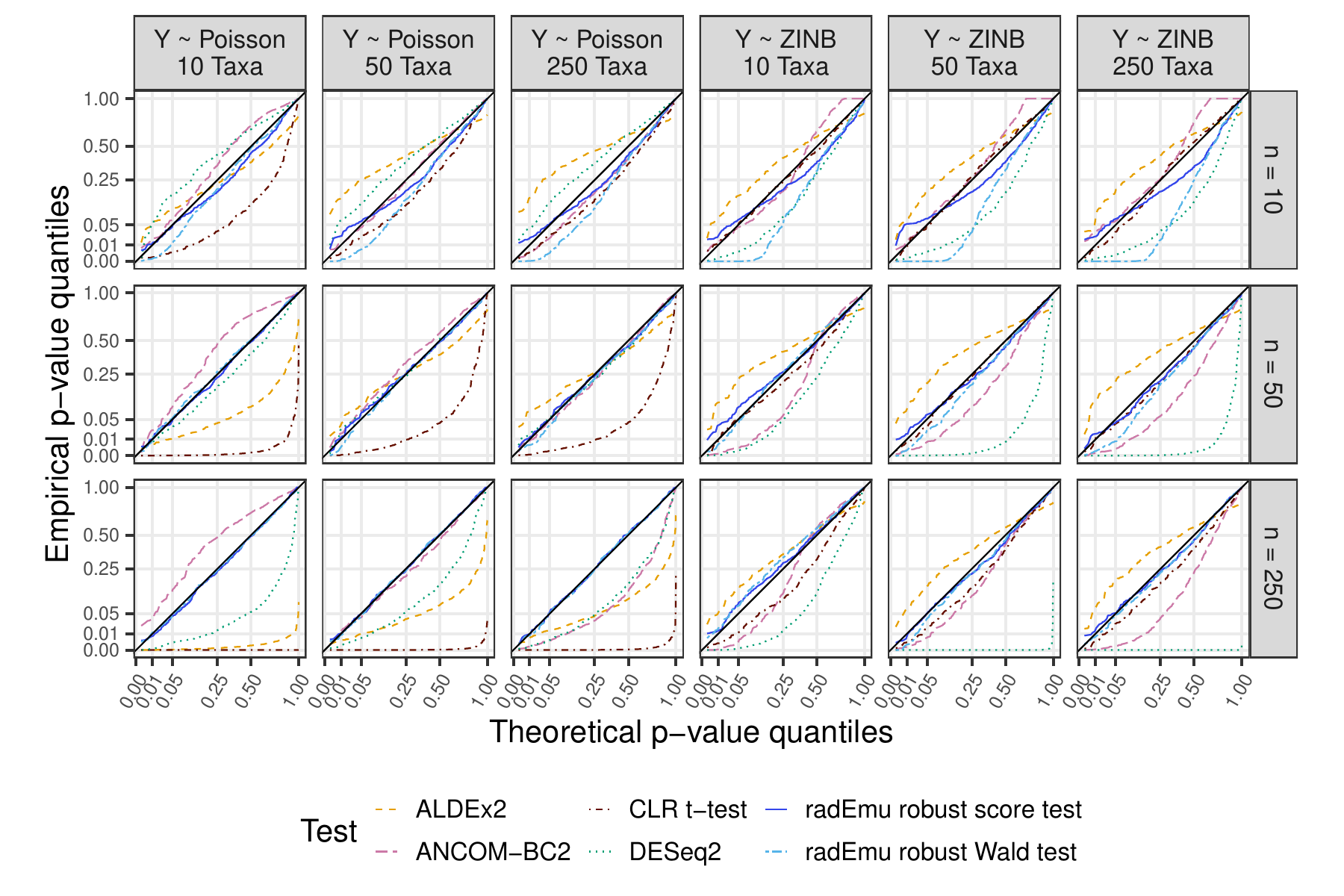}}
	\caption{Q-Q plots comparing empirical quantiles (y-axis) of the robust score (dark blue) and robust Wald (light blue) p-values to theoretical quantiles (x-axis). P-values from ALDEx2 (orange), ANCOM-BC2 (pink), a t-test on centered log ratio (CLR) transformed abundances (brown), and DESEq2 (green) are also compared.
	A conservative test will produce a curve above the line $x = y$ for small p-values and an anti-conservative test will produce
	a curve below it. To aid in comparison of lower quantiles, both axes are presented on a square-root scale. 500 simulations were performed.}  \label{fig:t1e}	\end{center}
	\end{figure} 

\begin{figure} 
	\begin{center}
	\centerline{\includegraphics[width=\textwidth]{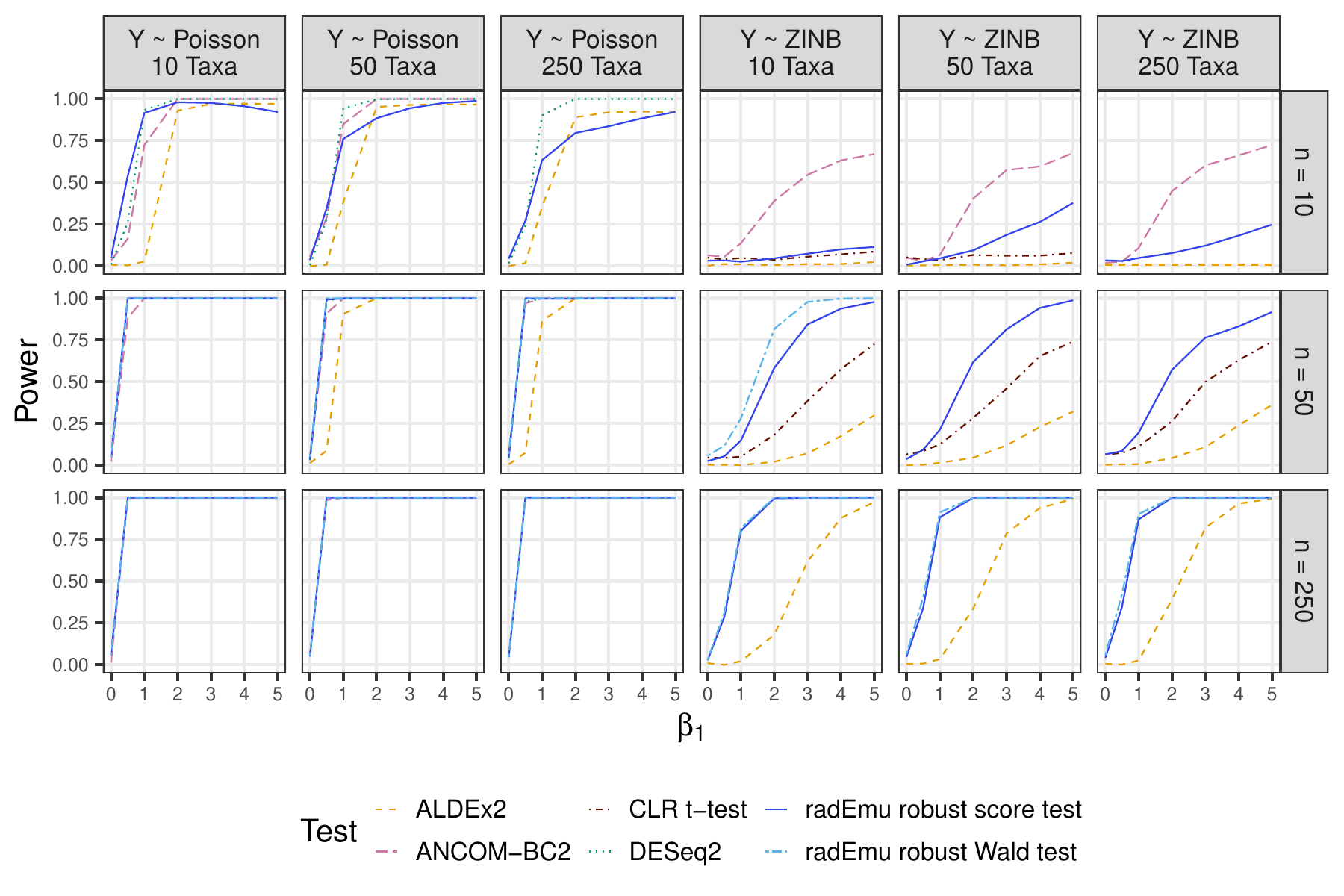}}
	\caption{Empirical power to reject $H_0$ at the 5\% level (y-axis) across a range of effect sizes (x-axis), estimated from 500 simulations. $\beta_1 = 1$ represents an $\approx$ 3-fold difference in the ratio of true abundances across groups while $\beta_1 = 5$ represents an $\approx$ 150-fold ratio. Only valid tests (those that control Type 1 error rates for the given sample size, number of categories and data distributions) are shown.}  \label{fig:power}	\end{center}
	\end{figure}

\section{Simulations} \label{sec:simulations}

We conduct a simulation study to examine the Type 1 error and power
of our proposed tests. We simulate in two distributional settings:
a correctly specified setting in which counts are independently Poisson-distributed, and a
misspecified setting in which the mean model is correctly specified but counts
follow independent zero-inflated negative binomial (ZINB) distributions.
To obtain draws from a ZINB distribution, we first draw from a Negative Binomial distribution
with mean $\mu$ given by our model and dispersion parameter $\phi = 5$ (thus, their variance is $\mu + \mu^2/\phi$). 
We multiply this draw with an independent Bernoulli draw $\xi$ with $Pr(\xi =1) = 0.4$ to obtain
our final ZINB draw.

In all
simulations, we simulate observations in two groups of equal size and test nulls of the form $\beta_{2j} - g_{p}(\beta_2) = 0$,
where $\beta_2$ corresponds to between-group differences in log mean (up to a constant shift).
 The pseudo-Huber
smoothing parameter for $g_{p}$ is 0.1 in all simulations. The first row of $\beta$ is fixed so that
its odd elements form an linearly increasing sequence from -3 to 3 and its even elements form a similarly decreasing
sequence from 3 to -3. We construct the first row of $\beta$ in this way so that it is not strongly correlated with the 
second row of $\beta$.  In null simulations, the second row of $\beta$, save its $J/2$- and $(J + 1)/2)$-th elements, which we set equal to zero, is given by $5\times\text{sinh}(x)/\text{sinh}(10)$ where
$x$ consists of linearly increasing sequence from $-10$ to $10$, and $\text{sinh}$ indicates the hyperbolic sine function. Under alternatives,
the second row of $\beta$ is the same as just described with the exception of its $J/2$-th element, 
which is set equal to a value from the set $\{0.5, 1, 2, 3, 4, 5\}$. 
The values of $\beta$ 
common to all hypotheses were chosen to reflect a setting in which most taxa vary similarly in mean concentration across
two groups, with a few taxa deviating from this typical variation. The values of $\beta_{k^{\star}j^{\star}}$ (i.e., the element of $\beta$ for which we 
we test equality to $g(\beta_{k^{\star}})$) are intended to reflect a range from a small ($\approx$ 1.7-fold deviation from 
typical ratio of mean concentrations across groups) effect, up to a strong ($\approx$ 150-fold deviation from typical) 
effect.
We simulate with total number of taxa $J \in \{10,50,250\}$ and
total sample size $n \in \{10,50,250\}$. Note that total sample
size indicates count of independent observations across both groups, so
$n = 10$ corresponds to $5$ observations in each of the two groups.

Figures \ref{fig:t1e} and \ref{fig:power} summarize the results of our simulations under the null and alternatives, respectively, with tabular results for the proposed methods in Supporting Information (Section \ref{si:sims}).
 As expected, the robust score test exhibits conservatism in simulations under the null, with greater conservatism at smaller sample sizes
and with ZINB-distributed counts. By contrast, the robust Wald test is anti-conservative in the smallest samples,
with empirical Type 1 error $0.316$ in ZINB
simulations and $0.160$ in Poisson simulations with $n = 10$, $J = 250$. In these
two simulation settings, robust score tests have empirical Type 1 error $0.034$ and $0.044$, respectively. Notably, while the robust Wald
test tends toward greater anti-conservatism when counts are not Poisson distributed, the robust score test appears to tend toward
greater conservatism. At the highest sample size we considered, $n = 250$, empirical Type 1 error is close to nominal regardless of
distribution, number of taxa, or proposed test. 

In simulations under the alternative, we observe higher power when counts are drawn from a Poisson than from a
ZINB distribution, and when the sample size is larger. While the robust Wald test exhibits higher power than the robust score test at smaller sample sizes, it also fails to control Type 1 error, so we do not recommend this test for use in small or moderate samples. 
We also observe a modest impact of total number of taxa $J$ on power: for ZINB simulations, we observe nondecreasing power in $J$. However,
power of the score test does appear to decrease in $J$ for Poisson simulations with $n = 10$.

We also compare error rates of our proposed methods to popular differential abundance methods that test comparable null hypotheses (Figures \ref{fig:t1e} and \ref{fig:power}). 
Three of the most widely-used methods (ALDEx2 \citep{fernandes2014unifying}, ANCOM-BC2 \citep{lin2024multigroup}, and a two-sample t-test on CLR-transformed counts) add a small non-zero ``pseudocount'' $\epsilon$ to all values prior to log-transformations. In part because $\log\left(\mathbb{E} Y_{ij}\right) \neq \mathbb{E} \left[\log \left(Y_{ij} + \epsilon \right)\right]$, these methods do not test identical null hypotheses to the proposed method. 
 However, the null hypothesis tested by these methods holds (for data generating mechanisms considered) when $\beta_1 = 0$, and does not hold when $\beta_1 \neq 0$. 
Accordingly, we treat as comparable the nulls tested by the proposed method, ALDEx2, ANCOM-BC2, a two-sample t-test on CLR-transformed counts, as well as DESeq2 \citep{love2014moderated}. The null hypothesis does not necessarily coincide for corncob \citep{martin2020modeling} or IFAA \citep{li2021ifaa}, and therefore we do not consider these methods in this comparison. 

In most settings investigated, ALDEx2 has conservative Type 1 error rates, resulting in its relatively low power compared to the proposed score test, though ALDEx2 can be anticonservative, with 5\% error rates as high as 28\% when $n = 250$, $J = 250$, and $Y$ follows a Poisson distribution. 
ANCOM-BC2 generally controls error rates when the observed counts follow a Poisson distribution, with the exception of $n = 250$ and $J = 250$, when its error rate at the 5\% level is 32\%. It also fails to control error rates in most ZINB-distributed count settings. 
Because methods that do not control error rates are not valid hypothesis tests ($\text{Pr}(p \leq \alpha) \leq \alpha$ under the null defines a hypothesis test), we only compare power amongst methods that control error rates. Accordinly, we find that ANCOM-BC2 has higher power than the proposed method for ZINB-distributed counts when $n = 10$. 
A two-sample t-test on CLR-transformed counts (with a pseudocount of 1 added to all $Y_{ij}$ values) controls error rates in most ZINB settings, but not in Poisson settings. When error rates are controlled, the CLR t-test has lower power than the proposed method. 
DESeq2 typically fails to control the Type 1 error rates, but has high power in the few cases where errors are controlled (Poisson data with $n = 10$). 
In contrast, the proposed score test controls error rates in all settings, and consistently displays high power. The power of the proposed score test can be outperformed (with respect to power) by the proposed Wald test and ANCOM-BC2, but these methods do not consistently control error rates. Overall, we find the proposed score test controls error rates across a wide variety of data generating processes, sample sizes, and species diversities; and high power to reject false null hypotheses when compared to rate-controlling alternatives.

\section{Data analysis: Colorectal cancer microbial metaanalysis} \label{sec:data}

We now investigate 
microbial strains that are unusually enriched or depleted
in populations with colorectal cancer (CRC)
compared to otherwise similar control populations. We consider a shotgun metagenomic dataset published by \citet{wirbel2019meta}, which is a meta-analysis of five studies (total $n = 575$), each of which obtained fecal samples from participants with
colorectal cancer and cancer-free controls. 
We consider the ``ideal'' but unobserved data $W_{ij}$ to be the cell concentration of microbial strain $j$ in fecal material of study participant $i$, and let $Y_{ij} \in \mathbb{R}_{\geq 0}$ be the observed abundance measured as the depth of coverage of marker genes calculated by the mOTU profiler v2 \citep{milanese2019microbial}.
2342 unique strains were detected. In addition to metagenomic data, \citet{wirbel2019meta} published clinical and demographic metadata including
colorectal cancer status, age in years,
binary gender, BMI, and timing of sampling relative to colonoscopy (sample taken
prior to or following colonoscopy).
\citet{wirbel2019meta} applied stringent filtering criteria to remove low-abundance strains from their dataset. 
While not necessary, to enable the comparison of our results with theirs, we applied the same filtering criterion (excluding all taxa that did not exceed empirical relative abundances of $10^{-3}$ in at least three studies). 
We excluded 9 participants with 
missing covariate data (8 for BMI and 1 for timing of sampling relative to colonoscopy). With these exclusions, we analyze $n =566$ study participants and $J=845$ metagenomic strains.


We fit our proposed model, applying Firth penalization as well as a smoothed median identifiability constraint to the rows of $\beta$. 
We include, in addition to an intercept and a
term for CRC status (1 for CRC and 0 otherwise),
terms in our model for age (linear spline with a single knot at median age 64 years),
BMI (linear spline with a single knot at median BMI 24.7 kg/m$^2$), gender (1 for Male and 0 otherwise),
study, and whether fecal samples were taken after colonoscopy (1 if after, 0 if prior). Thus, coefficients on CRC status estimate the difference in log mean cell concentration comparing
CRC patients to non-CRC controls who are alike in gender, BMI, age, study, and timing of stool sampling relative to colonoscopy, relative to typical such differences across taxa. We perform robust score tests and control the false discovery rate (using the \cite{benjamini1995controlling} (BH) procedure, following \cite{wirbel2019meta}) at $0.1$. 

\begin{figure}
	\begin{center}
	 \centerline{\includegraphics[width=\textwidth]{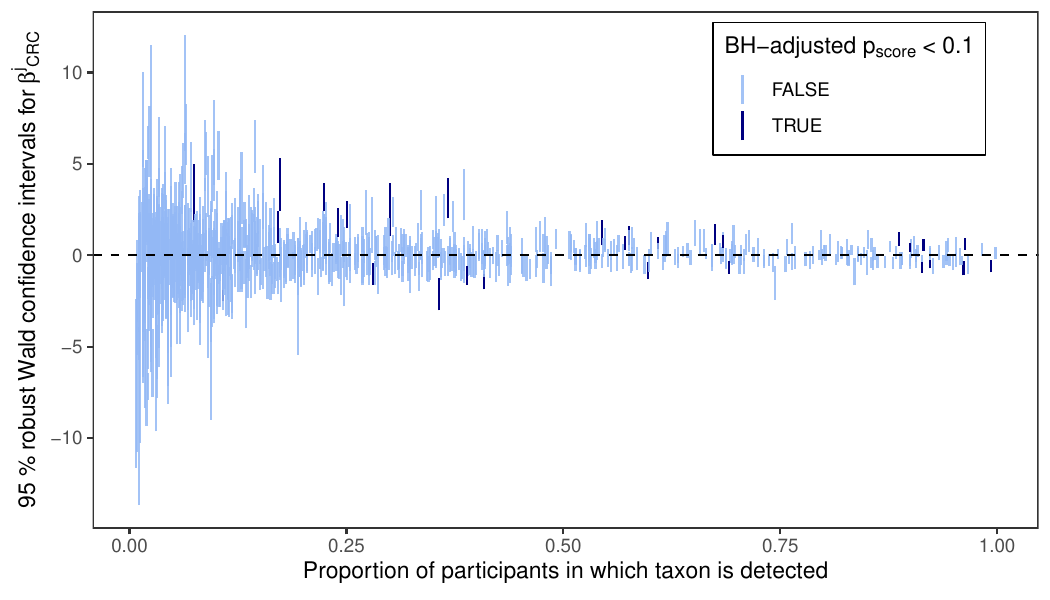}}   
	\caption{95\% Wald confidence intervals for taxon-specific CRC status effects estimated in our multivariable model (y-axis), plotted against the proportion of participants in all studies in which each
	taxon was detected (x-axis).  Taxa with Benjamini-Hochberg-adjusted robust score p-values below 0.1 are shown in dark blue, with all others shown in light blue. Significantly differentially abundant taxa are identified across levels of observed prevalence.}  \label{wirbel_fig1}
	\end{center}
	\end{figure}

\begin{figure}
	\begin{center}
	\centerline{\includegraphics[width=\textwidth]{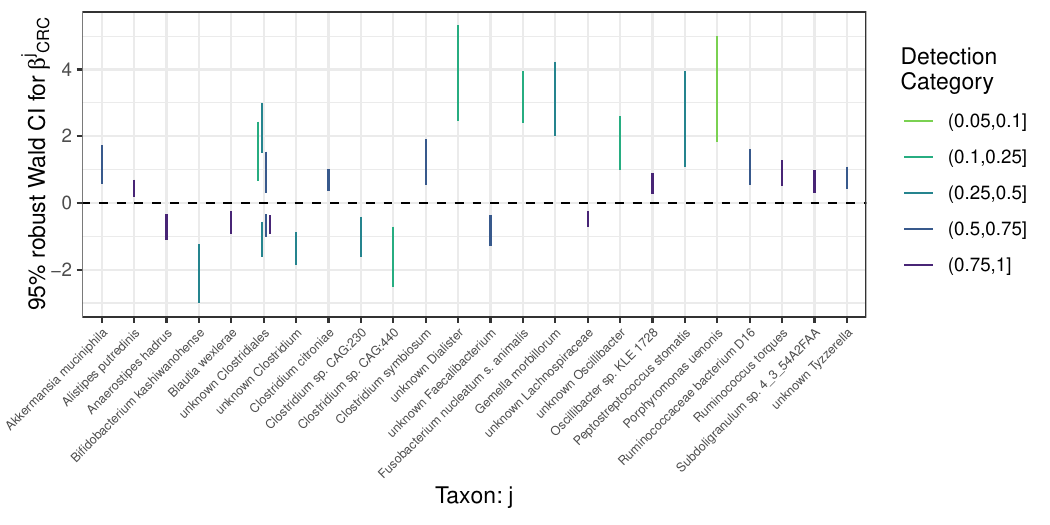}}
	\caption{Taxa associated with CRC status in adjusted model at FDR level 0.1. Species designations
	are given on the x-axis; multiple confidence intervals are reported for unknown Clostridiales as multiple mOTUs with no
	known species designation were identified as significantly associated with CRC status in this class. 
	95\% Wald confidence intervals for effect estimates (taxon-specific difference from smoothed median difference in log mean cell concentrations
	comparing CRC patients to controls, all other covariates held constant) are shown on the y-axis, with color indicating proportion of study participants in whom a given taxon was detected.}  \label{wirbel_fig2}
	\end{center} 
	\end{figure}

Robust Wald confidence intervals for the coefficients on CRC status are shown in Figure \ref{wirbel_fig1}. 
We identify 30 taxa for which we reject the null hypothesis of equal log mean cell concentration for CRC patients and (otherwise alike) non-CRC controls at a false discovery rate of 0.1 based on the robust score test. These taxa are denoted by navy-colored confidence intervals. 
Taxa are ordered along the x-axis by the proportion of participants in which they are detected, highlighting that significantly differentially abundant taxa are identified across the spectrum of detection levels. 
Taxa our analysis identifies as more enriched than typical among CRC patients relative
to similar controls include an unknown \textit{Dialister} species estimated to have ratio of mean cell concentration 49 times greater than typical
(95\% CI $12 - 210$; $p = 1.1\times10^{-6}$),
 \textit{Porphyromonas uenonis} (30 times greater; 95\% CI $6.2-150$; $p = 0.0025$),
\textit{Fusobacterium nucleatum s. animalis} (24 times greater; 95\% CI $11 - 52$; $p = 2.2\times10^{-5}$),
and 
\textit{Gemella morbillorum} (23 times greater; 95\% CI $7.6 - 69$; $p = 0.0016$). Confidence intervals for all effect sizes that are significantly 
associated with CRC status at FDR cutoff $0.1$ are given in Figure \ref{wirbel_fig2}.

We identify a number of taxa associated with CRC status that are detected in relatively high 
 proportion in the study populations. For example, we estimate for \textit{B. kashiwanohense}, which was detected in 35.7\% of study participants, 
 that the ratio of mean stool cell concentrations for this taxon comparing CRC patients to alike
 controls is 8.6 times lower (95\% CI 3.6 -- 20.9; unadjusted robust score $p = 0.0022$) than the typical such 
 ratio among taxa included in our analysis. This is consistent with a possible protective effect of \textit{B. kashiwanohense}, but 
 we cannot rule out association due to unmeasured confounding (e.g., through diet) or reverse causation (e.g., development of CRC 
 induces relative depletion of \textit{B. kashiwanohense}). We note that \textit{B. kashiwanohense} was not identified as a taxon negatively 
 associated with CRC status in 
 the analysis of \citet{wirbel2019meta}.


\begin{figure} 
	\begin{center}
		 \centerline{\includegraphics[width=\textwidth]{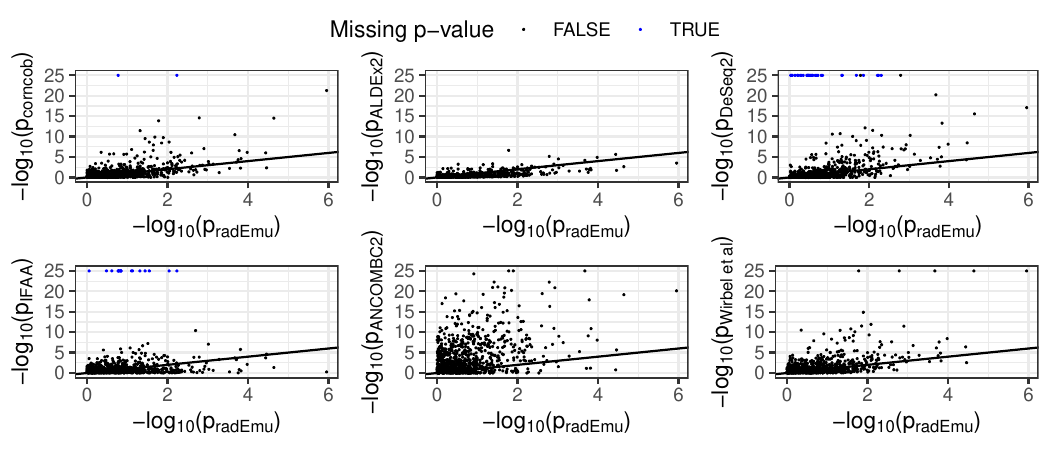}}
	\caption{Negative $\log_{10}$ p-values from distinct approaches to differential abundance analyses based on the \citet{wirbel2019meta} dataset. The proposed method's robust score test p-values are shown on the x-axis, and comparator methods on the y-axis. 
	\citet{wirbel2019meta} report some p-values equal to zero; to allow graphing on the log scale, we 
	replace these values with $10^{-25}$. 
	In addition to the missing/uncomputable p-values of corncob,  DESeq2 and IFAA, two DESeq2 p-values and three ANCOM-BC2 p-values were so small as to make plotting infeasible; these observations correspond to $(p_{\text{radEmu}}, p_{\text{DESeq2}})$ tuples of $(10^{-1.8}, 10^{-31})$ and $(10^{-2.8}, 10^{-33})$ and $(p_{\text{radEmu}}, p_{\text{ANCOM-BC2}})$ tuples of $(10^{-1.8}, 10^{-26})$, $(10^{-3.7}, 10^{-39})$, and $(10^{-1.9}, 10^{-26})$. } \label{p_compare}
\end{center}
\end{figure}

We compare our results to those of \citet{wirbel2019meta}, as well as to 5 other popular differential abundance methods in the microbiome literature, in Figure \ref{p_compare}. All panels show robust score p-values obtained from our adjusted model on the x-axis, and p-values obtained from comparator methods on the y-axis. 
We briefly describe details for each method here; see the Code Appendix for a reproducible workflow. 
(Top left) corncob \citep{martin2020modeling} likelihood ratio test p-values with design matrix as for the proposed method and timing of sampling and study as dispersion covariates; 
(top center) ALDEx2 \citep{fernandes2014unifying} with centered log-ratio transform and design matrix as for the proposed method and 512 Monte Carlo iterations; 
(top right) DESeq2 \citep{love2014moderated} likelihood ratio test with design matrix as for the proposed method and ``local'' estimation of dispersions; 
(lower left) IFAA \citep{li2021ifaa} with design matrix as for the proposed method; 
(lower center) ANCOM-BC2 \citep{lin2024multigroup} with design matrix as for the proposed method and without prevalence filtering; and 
(lower right) the results of 
\citet{wirbel2019meta}, who report a blocked Wilcoxon 
test performed on proportion-scale mOTU data, with blocking on study and timing of 
sampling relative to colonoscopy (five p-values of zero are altered to $10^{-25}$ for visual comparison). 
The proposed method produces smaller p-values in 44\%, 68\%, 52\%, 49\%, 20\% and 41\% of cases for corncob, ALDEx2, DESeq2, IFAA, ANCOM-BC2 and blocked Wilcoxon respectively. 
Thus, consistent with the results of our simulation study (ZINB, $J= 250$ and $n=250$), our method typically produces smaller p-values than ALDEx2 (which was generally found to be conservative) but larger p-values than ANCOM-BC2 (which was generally found to be anti-conservative). 
In general, we observe weak correlations between all methods' p-values, consistent with documented discrepancies between widely used differential abundance methods in microbiome science \citep{nearing2022microbiome}. That said, our semiparametric estimation procedure, model-robust inference method, and simulation study results give confidence that our proposed method presents reliable inference with high power. 

We also contrast our results with those of \citet{wirbel2019meta}, finding substantial variation in the ordering of p-values between our proposed method and the blocked Wilcoxon test (Spearman correlation $0.44$). Only 45 of the 94 taxa \citet{wirbel2019meta} identified as
significantly associated with CRC status are among the 94 taxa most significantly associated with CRC status in our analysis, 
and of the 30 taxa we highlight at FDR level $0.1$, only 10 are among the 30 most highly significant taxa identified by \citet{wirbel2019meta}. We believe that this lack of concordance is likely due to two major factors.
Firstly, our model adjusts for a larger set of potential confounders than do \citet{wirbel2019meta}, which may account for our discrepant findings. 
In addition, we estimate effects on the scale of log mean cell concentration, whereas \citet{wirbel2019meta} conduct tests on the read proportion scale. Because a perturbation to the abundance of a small number of taxa on the cell concentration scale induces a dense perturbation on the proportion scale, analyses on the proportion scale may highlight taxa that do not vary in a scientifically interesting way on the cell concentration scale. Similarly, analyses performed on proportion data 
may fail to detect scientifically meaningful effects on the cell concentration scale because of fluctuations in cell concentration among other taxa. 

While it is beyond the scope of the present study to experimentally validate which taxa are truly differentially abundant in colorectal cancer gut microbiomes, we investigate the empirical number of false discoveries of all methods applied to this dataset using a permutation approach. We create 20 datasets $(\mathbf{X}^{*}, \mathbf{Y}^{*})$ with no association between microbial abundances and cancer diagnosis by randomly reordering the rows of the design matrix $\mathbf{X}$ such that the permuted design matrices $\mathbf{X}^*$ satisfies $\text{Pr}(X_i^{*} = X_{i'}) = 1/n$ for all $i, i'$, but $\mathbf{Y} = \mathbf{Y}^*$. This approach leaves the joint distribution of the covariates unchanged but guarantees conditional independence between $Y_{i \cdot}^*$ and $X_{i, \text{CRC}}^*$, thereby guaranteeing that $\beta_{\text{CRC},j}^* = 0$ for all $j$. We ran all analyses as before on the permuted 20 datasets, and report the distribution of the number of taxa that each method determines to be differentially abundant when controlling the false discovery rates at 10\% (the empirical false discovery rate) in SI Table 3. 
Remarkably, our proposed method (radEmu score test) and ALDEx2 make no false discoveries across any of the 20 datasets. 
ANCOM-BC2 has the highest discovery proportion in this null dataset (median 37\%; IQR: 36--39\%), followed by the radEmu Wald test (which we do not recommend in general; median 3.2\%; IQR: 2.6--3.6\%), DESeq2 (median 0.5\%; IQR: 0.2--1.1\%), IFAA (median 0\%; IQR: 0--0\%; max 0.2\%), and the CLR t-test (median 0\%; IQR: 0--0\%; max 0.1\%). 
These results align with our simulation study, which demonstrated consistent error rate control of the proposed score test, very conservative behavior of ALDEx2, and anti-conservative behavior of the t-test, DESeq2 and ANCOM-BC2 for $n = 250, J = 250$ and zero-inflated Negative Binomially-distributed data. That said, we found DESeq2 to make fewer false discoveries than ANCOM-BC2, unlike in the the simulation study, which may be attributable to the differences between the ZINB distribution and the (unknown) true data generating process. 
ANCOM-BC2 also reports if discoveries pass a ``sensitivity score filter'', but even with this sensitivity filter, a median of 14 discoveries were made (median 1.6\%; IQR: 1.3--2.0\%; max 3.3\%). 
Combining these results with the findings of Section \ref{sec:simulations}, we find DESeq2 and ANCOM-BC2 to be anticonservative in general (consistent with the benchmarking study of \cite{cho2023comprehensive}), and similarly for the CLR t-test but to a lesser degree. We find the performance of the proposed method to be most similar to that of ALDEx2, with the proposed method having higher power than ALDEx2 without sacrificing error rate control.

\section{Discussion} \label{sec:discussion}

In this paper we present a regression method for inference on means of a nonnegative
multivariate outcome $W$ observed after perturbation by unknown
sample-specific scaling terms $z$
and category-specific detection effects $\delta$. Our method is motivated by microbiome sequencing experiments, in which outcomes of interest (microbial cell concentration across many species)
are typically observed indirectly via sequencing, which is known to be subject to
both sample-specific scaling and taxon-specific
distortion. However, we anticipate that our method may find
application in other settings, especially in other 'omics fields where similar scaling and detection effects are likely to distort measurements.

Our method has a number of advantages over existing methods commonly applied to perform differential abundance analyses in microbiome studies. Firstly, it targets a population estimand that can be distinguished from sample- and category-scaling effects and that exists under realistic conditions (positivity of population mean abundances $\mathbb{E}W_{\cdot j}$). Notably, we do not require positivity of true abundances $W_{ij}$ nor observed abundances $Y_{ij}$. Secondly, our method does not require data transformations, such as log-ratio transformations or relative abundance transformations. This allows us to use information regarding the precision of observations in estimation. It also avoids ad-hoc data manipulation (e.g., adding a small pseudocount to zero observations, restricting attention to nonzero counts, or infering whether observed zeroes are due to insufficient sequencing depth) that is necessary to apply log-ratio transformations to data containing zeroes. 
Additionally, we implement a flexible inference approach  that controls Type 1 error even in small samples with large excess-Poisson dispersion and high degrees of sparsity, and observed that the approach made no false discoveries on a shotgun sequencing dataset with true signal removed. 
In a simulation study, the proposed score test was the only method with error rate control in all data settings, and had comparable power to other error rate-controlling methods in most of these settings.
Finally, we incorporate an identifiability constraint that may be flexibly specified to suit varying scientific settings. In particular, we do not 
require a reference taxon (though if available, the reference taxon can define the identifiability constraint; see Example 1) nor the assumption of a sparse signal (see Example 2).

Our method is not without limitations, and admits several possible extensions. 
While our target estimand is identifiable in the presence of systematic measurement error in the form of multiplicative row and column perturbations to its mean, no such guarantees
exist for other forms of measurement error. A notable form of measurement error that is not addressed by our method is cross-sample contamination. Accordingly, we encourage the use of decontamination techniques \citep{davis2018simple, clausen2022modeling} prior to the use of
our proposed method. In addition, in contexts where it is scientifically plausible that the parameter of interest
$\beta_{k}$ is sparse, enforcing sparsity via $\ell_1$ penalization is an attractive option.
However, we note that if most elements of $\beta_k$ are truly equal to zero, $\beta_k$ is correctly specified under pseudo-Huber constraints up to negligible differences between medians and pseudo-Huber centers (see Example 2).
Furthermore, while estimation will be valid in the presence of longitudinal or cluster-dependent observations, it remains to extend our inferential methods to this setting. 
Finally, computation can become prohibitively expensive when $J \gtrapprox 10^4,$ and exact and approximate algorithms could be explored to enhance the scalability of the method.









\section{SI: Proofs of identifiability and asymptotic results} \label{appA}

Here, we provide formal proofs of the results stated in the text.

\begin{lemma} \label{thm:identifiably}
  For arbitrary $b \in \mathbb{R}^{p\times J}$, define $$G_{b} = \{\beta^{\star} \in \mathbbm{R}^{p \times J}: \beta^{\star} = b + \alpha\mathbf{1}_J^T \text{ for some } \alpha \in \mathbb{R}^p\}.$$
  If $\mathbf{X}$ has full column rank, for any $\beta^{\star}, \beta^{\dagger} \in \mathbb{R}^{p\times J}$ there exist $\mathbf{z} \in \mathbb{R}^n$ and $\mathbf{z}^{\dagger} \in \mathbb{R}^n$ such that
  under model (3),
  $\log\mathbb{E}[\mathbf{Y}|\mathbf{X}, \beta^{\star}, \mathbf{z}] = \log\mathbb{E}[\mathbf{Y}|\mathbf{X}, \beta^{\dagger}, \mathbf{z}^{\dagger}]$ if and only if $\beta^{\dagger} \in G_{\beta}$.
  \end{lemma} 

\begin{proof}
Suppose $\beta \in G_{\beta}$. Then there exists an $\alpha \in \mathbb{R}^p$ such that $\beta = \beta^{\star} + \alpha\mathbf{1}_J^T$ and by ((4), main text), for any $\mathbf{z}$ and
$\mathbf{z}^{\star} : = \mathbf{z} + \mathbf{X}\alpha$, $\log\mathbb{E}[\mathbf{Y}|\mathbf{X}, \beta^{\star}, \mathbf{z}] = \log\mathbb{E}[\mathbf{Y}|\mathbf{X}, \beta, \mathbf{z}^{\star}]$.
Now suppose $\beta \not\in G_{\beta^{\star}}$. If there exist $\mathbf{z}, \mathbf{z}^{\star}$ such that $\log\mathbb{E}[\mathbf{Y}|\mathbf{X}, \beta^{\star}, \mathbf{z}] = \log\mathbb{E}[\mathbf{Y}|\mathbf{X}, \beta, \mathbf{z}^{\star}]$, we have
\begin{align}
&\mathbf{z}\mathbf{1}_J^T + \mathbf{X}\beta^{\star} = \mathbf{z}^{\star}\mathbf{1}_J^T + \mathbf{X}\beta \notag \\
\Rightarrow ~ & \mathbf{X}(\beta^{\star} - \beta)  = (\mathbf{z}^{\star} - \mathbf{z})\mathbf{1}_J^T \notag \\
\Rightarrow  ~ &\mathbf{X}^T\mathbf{X}(\beta^{\star} - \beta)  = \mathbf{X}^T(\mathbf{z}^{\star} - \mathbf{z})\mathbf{1}_J^T \notag \\
\Rightarrow  ~ &(\beta^{\star} - \beta)  = (\mathbf{X}^T\mathbf{X})^{-1}\mathbf{X}^T(\mathbf{z}^{\star} - \mathbf{z})\mathbf{1}_J^T \notag \\
\Rightarrow  ~&\beta = \beta^{\star} + \tilde{\alpha}\mathbf{1}_J^T \text{ for } \tilde{\alpha} : = (\mathbf{X}^T\mathbf{X})^{-1}\mathbf{X}^T(\mathbf{z}^{\star} - \mathbf{z}) \in \mathbb{R}^p \notag \\
\Rightarrow ~&\beta \in G_{\beta^{\star}} \notag 
\end{align}
which is a contradiction. Hence $\beta \not\in G_{\beta^{\star}}$ guarantees
the impossibility of $\log\mathbb{E}[\mathbf{Y}|\mathbf{X}, \beta^{\star}, \mathbf{z}]$ $ = \log\mathbb{E}[\mathbf{Y}|\mathbf{X}, \beta, \mathbf{z}^{\star}]$. 
\end{proof}

\begin{assumption} \label{assumption:mean}
	Let $(X, Y, Z) \overset{iid}{\sim} P^{(0)}$ for unknown true data generating mechanism $P^{(0)}$ with bounded support in $\mathbb{R}^p \times \mathbb{R}^J \times \mathbb{R}.$ Then $P^{(0)}$ satisfies $\log \mathbbm{E}[Y|X,Z;\beta^{(0)}] = Z\mathbf{1}_J + {\beta}^{(0)T}X$ for ${\beta}^{(0)} \in \mathbb{R}^{p \times J}$ with $J$-th column equal to $\mathbf{0}_J$. 
\end{assumption}

\begin{assumption}	\label{assumption:pd}
	$P^{(0)}$ satisfies $\Sigma_X := Cov(X)$ is positive definite. 
\end{assumption}

\begin{lemma} \label{lemma:maximizer}
	Under Assumptions \ref{assumption:mean} and \ref{assumption:pd}, $\beta^{(0)}$ is the unique maximizer of the population criterion $M(\beta) := \mathbbm{E}_{P^{(0)}}\hspace{0.05cm}l_i^{\text{profile}}(\beta)$. 
\end{lemma}

\begin{proof}
	The law of iterated expectation, applying $\log \mathbbm{E}[Y|X,Z;\tilde{\beta}^{(0)}] = Z\mathbf{1}_J + \tilde{X}\tilde{\beta}^{(0)}$, and factorizing gives
	\begin{align*}
	\mathbbm{E}&_{P^{(0)}}\hspace{0.05cm}l_i^{\text{profile}}(\beta) =  \mathbbm{E}\left[\mathbbm{E}_{Y|X,Z}\hspace{0.05cm}l_i^{\text{profile}}(\beta)\right] \\
	&=  \mathbbm{E}_{X,Z}\left[\mathbbm{E}_{Y|X,Z}\hspace{0.05cm} \sum_{j = 1}^J \left[ Y_j \log \left( \frac{\exp (1_{\{j<J\}}X^T\beta^j)}{1 + \sum_{j' = 1}^{J - 1}\exp(X^T\beta^{j'})} \right) - \big|\big|Y\big|\big|_1 \frac{\exp (1_{\{j<J\}}X^T\beta^j)}{1 + \sum_{j' = 1}^{J - 1} \exp(X^T\beta^{j'})}\right]\right] \\
	&=  \mathbbm{E}_{X,Z}\Big[\sum_{j = 1}^J \Big[ \exp\left(1_{\{j<J\}}X^T\beta^{(0) j} + Z\right) \log \left( \frac{\exp (1_{\{j<J\}}X^T\beta^j)}{1 + \sum_{j' = 1}^{J - 1}\exp(X^T\beta^{j'})} \right)\\
	&\hspace{3cm}- \left[ \sum_{j' = 1}^J \exp(1_{\{j'<J\}}X^T\beta^{(0) j'} + Z)\right] \frac{\exp (1_{\{j<J\}}X^T\beta^j)}{1 + \sum_{j' = 1}^{J - 1} \exp(X^T\beta^{j'})}\Big]\Big] \\
	&=  \mathbbm{E}_{X,Z}\exp(Z)\left(1 +  \sum_{j = 1}^{J - 1}\exp\left(X^T\beta^{(0) j} \right)\right) \times \\
	&\hspace{0.5cm} \sum_{j = 1}^J \left[ \frac{\exp(1_{\{j<J\}}X^T\beta^{(0) j})}{1 +  \sum_{j' = 1}^{J - 1}\exp(X\beta^{(0) j'} )} \log \left( \frac{\exp (1_{\{j<J\}}X^T\beta^j)}{1 + \sum_{j' = 1}^{J - 1}\exp(X^T\beta^{j'})}\right) -  \frac{\exp (1_{\{j<J\}}X^T\beta^j)}{1 + \sum_{j' = 1}^{J - 1} \exp(X^T\beta^{j'})}\right] 
	\end{align*}
	The sum in the interior of the final expression is of form $\mathbf{c}^T\log\mathbf{b} - \mathbf{1}^T\mathbf{b}$, which is a strictly convex function maximized at $\mathbf{b} = \mathbf{c}$. Therefore, since $\beta^{(0)}$ maximizes the conditional expectation of the penalized log likelihood at every value of $X$ and $Z$, it follows that $\beta^{(0)}$ is a maximizer of $M(\beta)$. Uniqueness of the maximizer follows from the (strict) convexity of $M(\beta)$. 
\end{proof}

\begin{assumption}  \label{assumption:bounded} 
  ${\beta}^{(0)}$ lies in a closed, bounded subset $\mathbf{\Omega}$ of $\mathbb{R}^{p\times J}$
  \end{assumption}
  
  \begin{theorem} \label{thm:consistency}
  Under Assumptions \ref{assumption:mean} -- \ref{assumption:bounded}, the maximum penalized likelihood estimator (MPLE) $\hat{\beta}^{(n)}$ is consistent. That is, $\hat{\beta}^{(n)}  \xrightarrow{\text{p}} \beta^{(0)}$. 
  \end{theorem}
  \begin{proof}
    We apply \citet[Theorem 5.7]{vanderVaart1998}. Let $m_\beta = m_\beta(x,y)$ denote the profile log likelihood evaluated at $\beta$, $x$, and $y$, and let 
  $M_n(\beta) := \frac{1}{n} \sum_{i = 1}^n m_{\beta}(X_i,Y_i)$. As the support is bounded (Assumption 1) and the parameter space is bounded (Assumption 3), the class $\{{m_{\beta}: \beta \in \Omega}\}$ attains a finite maximum over $\mathcal{X} \times \mathcal{Y} \times{\Omega}$ and is therefore $P^{(0)}$-Glivenko-Cantelli \cite[Chapter 5, p46]{vanderVaart1998}. Hence, $\sup_{\beta \in \Omega} |M_n(\beta) - M(\beta)| \xrightarrow{\text{p}} 0$. By Lemma \ref{lemma:maximizer} and the convexity of $M(\beta)$, the point of maximum $\beta^{(0)}$ is well-separated.
  
  It remains to show that the maximizer of the penalized likelihood satisfies $M_n(\hat{\beta}_n) \geq M_n(\beta^{(0)}) - o_p(1)$. 
  By (9) and determinant identities, we have 
  \begin{align*}
  \frac{1}{n} l_n^{\text{penalized}}(\beta) 
  & = \frac{1}{n}l_n^{\text{profile}}(\beta) + \frac{1}{2n}\left[\log \big|\frac{1}{n}\tilde{\mathbf{X}}^T\mathbf{V}(\beta)\tilde{\mathbf{X}}\big| + p(J-1)\log(n)\right],
  \end{align*}
  and by definition of $\hat{\beta}_n$ we have $\frac{1}{n} l_n^{\text{penalized}}(\hat{\beta}_n) \geq \frac{1}{n} l_n^{\text{penalized}}(\beta)$ for any $\beta$. Therefore
  \begin{align*}
  M_n(\hat{\beta}_n)  &\geq M_n(\beta^{(0)}) 
  - \frac{1}{2n}\left[\log \big|\frac{1}{n}\tilde{\mathbf{X}}^T\mathbf{V}(\hat{\beta}_n)\tilde{\mathbf{X}}\big| - \log \big|\frac{1}{n}\tilde{\mathbf{X}}^T\mathbf{V}(\beta^{(0)})\tilde{\mathbf{X}}\big| \right].
  \end{align*}
  Assumptions 2, 3, and the law of large numbers give that $\frac{1}{n} \tilde{\mathbf{X}}^T\mathbf{V}(\beta)\tilde{\mathbf{X}}$ has a stable limit in probability, which we denote $I_{\beta}$. Thus $M_n(\hat{\beta}_n) \geq M_n(\beta^{(0)}) - o(\frac{1}{n})o_p(1)$. The result follows.
  \end{proof}

\begin{theorem} \label{thm:normality}
  Under Assumptions \ref{assumption:mean} -- \ref{assumption:bounded}, the MPLE $\sqrt{n}\left(\hat{\tilde{\beta}}^{(n)} - \tilde{\beta}^{(0)}\right) \overset{\mathcal{D}}{\rightarrow} \mathcal{N}(\mathbf{0},\Sigma)$ for positive definite $\Sigma$. 
  \end{theorem}
  \begin{proof}
  For vector-valued $\tilde{\beta} \in \mathbb{R}^{p(J - 1)}$, we have differentiability of $m_{\tilde{\beta}}(x,y)$ as well as the existence of a nonsingular second 
  derivative $I_{\tilde{\beta}}$ of the population criterion $M(\tilde{\beta})$. 
  We previously showed (Theorem 1 and proof) that $M_n(\hat{\beta}_n) \geq M_n(\beta^{(0)}) - o_p(\frac{1}{n})$ and $\hat{\tilde{\beta}} \xrightarrow{p} \tilde{\beta}^{(0)}$. Therefore we may apply \citet[Theorem 5.23]{vanderVaart1998} to obtain the desired result, noting that $\Sigma = I_{\tilde{\beta}_0}^{-1} \left( \mathbb{E} [ \nabla m_{\tilde{\beta}} \nabla^T m_{\tilde{\beta}}]\big|_{\tilde{\beta} = \tilde{\beta}_0} \right) I_{\tilde{\beta}_0}^{-1}$. 
  \end{proof}

\begin{theorem} \label{thm:normality_g}
  Fix $k \in \{1, \ldots, p\}$ and $j \in \{1, \ldots, J\}$. 
  Define $\hat{\gamma}^{(n)} := \hat{\beta}_{kj}^{(n)} - g(\hat{\beta}_k^{(n)})$ for a smooth function $g: \mathbb{R}^{J} \rightarrow \mathbb{R}$.  
  Under Assumptions \ref{assumption:mean} -- \ref{assumption:bounded}, $\hat{\gamma}^{(n)} \xrightarrow{\text{p}} \beta^{(0)}_{kj} - g(\beta^{(0)}_{k})$, and $\sqrt{n}\left(\hat{\gamma}^{(n)} - (\beta^{(0)}_{kj} - g(\beta^{(0)}_{k}))\right) \overset{\mathcal{D}}{\rightarrow} N\left(0, H^T\Sigma H\right)$ for some matrix $H$.
  \end{theorem}
  \begin{proof}
    Consistency follows from the continuous mapping theorem applying the function $h(\mathbf{b}) = b_{j} - g(\mathbf{b})$ for $h: \mathcal{B} \rightarrow \mathbb{R}$ where $\mathcal{B}$ is the set of $J$-dimensional vectors with $J$-th entry equal to zero. Convergence in distribution follows from the delta method. The matrix $H$ can be constructed by considering the derivative of the appropriate generalization of $h$ to vector-valued arguments in $\mathbb{R}^{p(J -1) \times 1}.$ 
  \end{proof}

  \begin{theorem} \label{thm:efficiency}
    Under Assumptions \ref{assumption:mean} -- \ref{assumption:bounded} and either $\text{Var}(Y| X,Z;\beta^{(0)}) = \phi \times$ $\text{diag} \left\{ \mathbbm{E}[Y|X,Z;\beta^{(0)}]\right\}$ or $\text{Var}(Y| X,Z;\beta^{(0)}) = \phi \tilde{\mathbf{V}}$ for unknown constant $\phi$, $\hat{\beta}_n$ is asymptotically efficient among asymptotically linear estimators of $\beta^{(0)}$. 
  \end{theorem}
  
  \begin{proof}
    By \cite{mccullagh1983quasi}, an estimator is optimal among linear estimators if it solves the quasi-likelihood equations with $\mathbbm{E}[Y] = \mu(\beta)$ and $\text{Var}[Y] = \phi V(\mu),$ assuming only these moment conditions. The Poisson MLE gives the solution under $\text{Var}(Y| X,Z;\beta^{(0)}) = \phi \times \text{diag} \left\{ \mathbbm{E}[Y|X,Z;\beta^{(0)}]\right\}$ and the Multinomial MLE gives the solution under $\text{Var}(Y| X,Z;\beta^{(0)})$ $= \phi \tilde{\mathbf{V}}$. We have shown that $\hat{\beta}^{\text{Poisson MLE}} = \hat{\beta}^{\text{Multinomial MLE}}$ when both quantities are well defined, and so by Theorem \ref{thm:consistency} and Slutsky, $\hat{\beta}_n - \hat{\beta}^{\text{Multinomial MLE}} = \hat{\beta}_n  - \hat{\beta}^{\text{Poisson MLE}} = o_p(1)$. Furthermore, $\lim_n \sqrt{n}\text{Var}(\hat{\beta}_n) = \lim_n \sqrt{n} \text{Var}(\hat{\beta}^{\text{Multinomial MLE}})$. Therefore, $\hat{\beta}_n$ attains the smallest asymptotic variance under $\text{Var}(Y| X,Z;\beta^{(0)}) = \phi \tilde{\mathbf{V}}$. But since $\hat{\beta}^{\text{Poisson MLE}} = \hat{\beta}^{\text{Multinomial MLE}}$ and $\hat{\beta}^{\text{Poisson MLE}}$ attains the smallest asymptotic variance under $\text{Var}(Y| X,Z;\beta^{(0)}) = \phi \times$ $\text{diag} \left\{ \mathbbm{E}[Y|X,Z;\beta^{(0)}]\right\}$, $\hat{\beta}_n$ must attain this variance bound also.
  \end{proof}

\begin{figure}
	\begin{center}
	\centerline{\includegraphics[width=\textwidth]{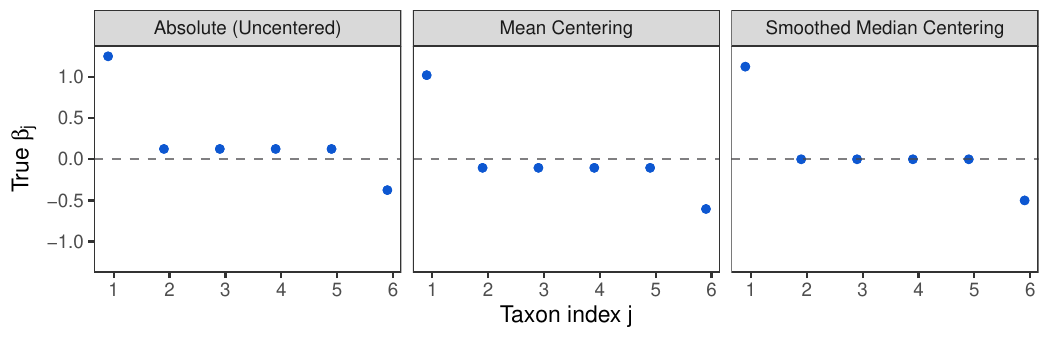}} 
	\caption{$\beta$ is only partially identifiable and cannot be estimated on the ``absolute'' scale without further assumptions (e.g., a reference category that is known to be equal in mean abundance across groups). Full identifiability is established via a constraint function, allowing us to estimate the log-fold differences in true abundances across groups relative to typical differences. The above illustrates the favorable qualities of centering via the pseudo-Huber smoothed median (right) compared to centering by the mean (middle). $\beta_j$ for $j=2, 3, 4, 5$ are unlikely to be of scientific interest (relative to $j=1, 6$), and the smoothed median constrains these ``uninteresting'' taxa closer to zero than mean centering. Note that centering by a smoothed median does not require any assumption of sparsity of true effects $\beta$ (left) to detect taxa changing in abundance.}  \label{fig:compare_centering}
	\end{center}
	\end{figure}

\section{SI: Pseudo-Huber Centering}

For a generic vector $x = [x_1, \dots, x_J] \in \mathbb{R}^J$ and smoothing parameter $\delta>0$, we define the pseudo-Huber smoothed median $c^{\star}$
of $x$ as follows:
\begin{align*}
c^{\star} &= \text{argmin}_c \sum_j \delta^2 \Big[ 1 + (\frac{x_j - c}{\delta})^2\Big]^{1/2} 
\end{align*}

We now derive the form of $\frac{\partial}{\partial x_j}c^{\star}$, which is needed for optimization and testing.

First, we have
\begin{align*}
&\frac{\partial}{\partial c}\Big[ \sum_j \delta^2 \Big[ 1 + (\frac{x_j - c}{\delta})^2\Big]^{1/2}\Big] |_{c = c^{\star}}  = 0 \\
\Rightarrow ~& \sum_j \frac{\delta^2}{2}\Big[ 1 + (\frac{x_j - c}{\delta})^2\Big]^{-1/2} 2(\frac{x_j - c}{\delta})(\frac{-1}{\delta}) = 0 \\
\Rightarrow ~& \sum_j w_j (x_j - c) = 0 \text{ for } w_j := \Big[ 1 + (\frac{x_j - c}{\delta})^2\Big]^{-1/2}
\end{align*}

We also have

\begin{align*}
\frac{\partial}{\partial x_{j^{\dagger}}} w_j & = -\frac{1}{2}\Big[ 1 + (\frac{x_j - c}{\delta})^2\Big]^{-3/2}2 (\frac{x_j - c}{\delta})(\frac{\mathbf{1}_{[j = j^{\dagger}]}-\frac{\partial}{\partial x_{j^{\dagger}}}c}{\delta}) \\
& = -w_j^3(\frac{x_j - c}{\delta^2})\Big[ \mathbf{1}_{[j = j^{\dagger}]}-\frac{\partial}{\partial x_{j^{\dagger}}}c\Big] \\
\end{align*}

Before proceeding, we will derive a useful identity. Let $ y := \frac{x_j - c}{\delta}$. Then we have

\begin{align*}
w_j - w_j^3 (\frac{x_{j} - c}{\delta})^2 & = (1 + y^2)^{-1/2} - (1 + y^2)^{-3/2}y^2 \\
& = \frac{1}{(1 + y^2)^{1/2}} - \frac{y^2}{(1 + y^2)^{3/2}} \\
& = \frac{1 + y^2}{(1 + y^2)^{3/2}} - \frac{y^2}{(1 + y^2)^{3/2}} \\
& = \frac{1}{(1 + y^2)^{3/2}} \\
& = w_j^3 
\end{align*}

We can now use implicit differentiation to find $\frac{\partial}{\partial x_j} c^{\star}$
\begin{align*}
&\frac{\partial}{\partial x_j} \sum_j w_j(x_j - c^{\star})  \\
=& \sum_j \Bigg[ w_j ( \mathbf{1}_{[j = j^{\dagger}]}-\frac{\partial}{\partial x_{j^{\dagger}}}c^{\star}) -w_j^3(\frac{x_j - c^{\star}}{\delta^2})\Big[ \mathbf{1}_{[j = j^{\dagger}]}-\frac{\partial}{\partial x_{j^{\dagger}}}c^{\star} \Big]\Bigg] =0 \\
\Rightarrow& \sum_j \Big[ w_j - w_j^3 (\frac{x_j - c^{\star}}{\delta})^2\Big]\frac{\partial}{\partial x_{j^{\dagger}}}c^{\star} = w_{j^{\dagger}} - w_{j^{\dagger}}^3 (\frac{x_{j^{\dagger}} - c^{\star}}{\delta})^2 \\
\Rightarrow & \sum_j w_j^3\frac{\partial}{\partial x_{j^{\dagger}}}c^{\star} = w_{j^{\dagger}}^3 \text{ by the identity above } \\
\Rightarrow & \frac{\partial}{\partial x_{j^{\dagger}}}c^{\star} = \frac{w_{j^{\dagger}}^3}{\sum_j w_j^3}
\end{align*}

For a given $x$, to find the pseudo-Huber center $c^{\star}(x)= \text{arg min}_c \sum_j \delta^2 \Big[ 1 + (\frac{x_j - c}{\delta})^2\Big]^{1/2}$ with parameter $\delta$ by minimizing successive quadratic approximations to $f(c;x,\delta) := \sum_j \delta^2 \Big[ 1 + (\frac{x_j - c}{\delta})^2\Big]^{1/2}$. For current value $c^{(t)}$ of $c$, we compute, for $j = 1, \dots, J$, $w_j^{(t)} := \Big[1 + (\frac{x_j - c^{(t)}}{\delta})^2\Big]^{-1/2}$. Letting $h^{(t)}(c) =  \sum_j \frac{1}{2}w_j^{(t)} (\frac{x_j - c}{\delta})^2$, we then construct quadratic approximation $\tilde{f}^{(t)} := f(c^{(t)})  - h^{(t)}(c^{(t)}) + h^{(t)}(c)$ to $f$ at $c^{(t)}$. We update $c^{(t + 1)}$ as the minimizer of $\tilde{f}^{(t)}$, which is 
conveniently the weighted mean of $x$ with weights $\frac{1}{\sum_j w_j}(w_1,\dots,w_J)$.

\section{SI: Algorithms and additional details regarding estimation} \label{weighted_supp}

\begin{algorithm}[H]
  \caption{Coordinate Descent for Unconstrained Maximum Likelihood Estimation} \label{ml_alg}
  \begin{enumerate}
  \item Initiate with
      \begin{itemize}
      \item Data $(X,Y)$
      \item Starting value $\beta^{(0)}$ for $\beta$
      \item Identifiability constraint $g(\beta_k) = 0$ for $k = 1,\dots, p$
      \item Convergence tolerance $\epsilon > 0$
      \item Iteration limit $t_{\text{max}}$
      \item Choice of working identifiability constraint $\beta^{j^{\dagger}} = 0$ (e.g., $j^{\dagger}$ such that $\sum_i \mathbf{1}_{Y_{ij^{\dagger}}>0}$ is maximized)
      \end{itemize}
  \item For $k = 1, \dots, p$, impose working constraint by setting $\beta_k = \beta_k - \beta_{kj^{\dagger}}$
  \item For iteration $t = 1,\dots, t_{\text{max}}$
    \begin{enumerate}
    \item For $j \in \{1,\dots,J\}/\{j^{\dagger}\}$, update $\beta^j$, the $j$-th column of $\beta$ with single iteration of algorithm (\ref{ml_score})
    \item Update $z_i$ for $i = 1, \dots, n$ via equation (7).
    \item If $\text{max}_{k,j} |\beta^{(t)}_{kj} - \beta^{(t - 1)}_{kj}| < \epsilon$ or 
    if $t = t_{\text{max}}$
      \begin{itemize}
      \item For $k = 1, \dots, p$ enforce original identifiability constraint by setting $\beta_k = \beta_k^{(t)} - g(\beta_k^{(t)})$.
      \item Exit algorithm and return $\beta$.
      \end{itemize}
    \item Otherwise increment $t$ by $1$ and return to step 3a.
    \end{enumerate} 
  
  \end{enumerate}
  \end{algorithm}

\begin{algorithm}[H]
  \caption{Scoring Update with Line Search (unconstrained)} \label{ml_score}
  \begin{enumerate}
  \item Initiate with
      \begin{itemize}
      \item Data $(X,Y^j)$
      \item Current values of $\beta^j$ and $\mathbf{z}$
      \item Maximum step size $s$
      \end{itemize}
  \item Then
    \begin{enumerate}
      \item Compute 
        \begin{itemize}
        \item gradient $S^j = X^T(Y^j - \boldsymbol{\mu}^j))$ for
      $\boldsymbol{\mu^j} = [\mu_1^j,\dots, \mu_n^j]^T = [\exp(X_1^T\beta^j + z_1),\dots, \exp(X_n^T\beta^j + z_n)]^T$
        \item  information $I^j = X^TWX$ for $W$ a diagonal matrix with $i$-th diagonal entry $\mu^j_i$ 
        \end{itemize}
      \item Compute update direction $d^j = {I^j}^{-1}S^j$. If $I^j$ is poorly conditioned, attempt to compute update direction as  $(I^j + \sigma\mathbf{I})^{-1}S^j$, sequentially increasing $\sigma$ until $(I^j + \sigma\mathbf{I})^{-1}$ is numerically invertible. ($I^j$ is theoretically always full rank if $X$ is full rank; in practice, it may be poorly conditioned numerically.)
      \item Perform line search to find update $\beta^j + \alpha d$ satisfying $l(\beta^j + \alpha d) \geq l(\beta^j) + c\alpha d^T\nabla l$ (i.e., the Armijo rule) where $l$ is the log likelihood function evaluated in $\beta^j$ with $z$ held fixed, 
      $c >0$ is a constant (by default we use $c = 10^{-4}$ as suggested by \citet{jorge2006numerical} in Chapter 17), $\nabla l$ is the partial derivative of the log likelihood with respect to $\beta^j$, and $\alpha$ is initiated at $0.5$ and 
      halved until the Armijo condition is met. After the Armijo condition is met, optionally rescale $ \alpha d$ so that $\max_k |\alpha d_k| \leq \xi$ for some maximum step length $\xi$ before accepting update  $\beta^j + \alpha d$ to prevent large steps that may lead to ill conditioning. We by default use $\xi = 1$.
          \end{enumerate}
  \end{enumerate}
  \end{algorithm}

\begin{algorithm}[H]
  \caption{Penalized Likelihood Estimation via Unconstrained Iterative Maximum Likelihood} \label{pl_alg}
  \begin{enumerate}
  \item Initiate with
      \begin{itemize}
      \item Data $(X,Y)$
      \item Starting value $\beta^{(0)}$ for $\beta$
      \item Identifiability constraint $g(\beta_k) = 0$ for $k = 1,\dots, p$
      \item Convergence tolerance $\epsilon > 0$
      \item Iteration limit $t_{\text{max}}$
      \end{itemize}
  \item Set $Y^{(1)} = \mathbf{Y} + \delta$ for some $\delta \geq 0$. (Choosing $\delta >0$ is not necessary but can ameliorate slow convergence issues that otherwise might arise in the 
  first iteration due to separation.)
  \item For iteration $t = 1,\dots, t_{\text{max}}$
    \begin{enumerate}
    \item Update $\beta^{(t)}$ using algorithm \ref{ml_score} applied to data $\mathbf{Y}^{(t)}$ with other arguments as specified above
    \item Update $\mathbf{Y}^{(t +1)}$ using $\beta^{(t)}$ and $z$ as described in Section 3 of main text
    \item If $\text{max}_{k,j} |\beta_{kj}^{(t)} - \beta_{kj}^{(t -1)}|< \epsilon$ or if $t = t_{\text{max}}$, exit and return $\beta$. Otherwise increment $t$ and return to step 3a.
  \end{enumerate} 
  \end{enumerate}
  \end{algorithm}
  
  
  \begin{algorithm}[H]
  \caption{Constrained Estimation} \label{constrained_algo}
  \begin{enumerate} 
    \item Initiate with
      \begin{itemize}
      \item Covariate matrix $\mathbf{X}$
      \item Initial value of $\beta$ and augmented outcomes $\tilde{\mathbf{Y}}$ obtained via fitting an unconstrained penalized model 
      \item Constraint function $g$
      \item Row and column indexes $k^{\star}, j^{\star}$ for the element of $\boldsymbol{\beta}$ to be constrained
      \item Initial value of augmented Lagrangian penalty parameter $\rho$
      \item Values $\kappa \in (0,1)$ and $\tau>1$ controlling scaling of $\rho$ in each iteration
      \item Convergence tolerance $\epsilon$
      \item Constraint tolerance $\gamma$
      \item Iteration limits $t_{max}$ for outer loop and $t^{inner}_{max}$ for inner loop
      \end{itemize}
    \item Set  $u^{(1)} = \rho(\beta_{k^{\star}j^{\star}} - g(\beta_{k^{\star}}))$ and $\rho^{(1)} = \rho$
    \item Then for iteration $t = 1, \dots, t_{max}$
    \begin{enumerate}
      \item Define $\mathcal{L}^{(t)} = -l(\beta,z) + u^{(t)}\big[\beta_{k^{\star}j^{\star}} - g(\beta_{k^{\star}})\big] + (\rho^{(t)}/2)\big[\beta_{k^{\star}j^{\star}} - g(\beta_{k^{\star}})\big]^2$
      \item Via block coordinate descent in $\beta$ and $z$, obtain $(\beta^{(t)},z^{(t)})$ satisfying $||\nabla \mathcal{L}^{(t)}||_2 < \epsilon$
        \begin{itemize}
          \item Coordinate descent steps in $\beta$ are approximate Newton steps using $-\nabla^2l + \rho \nabla g \nabla^T g$ as an approximation to the Hessian of the augmented Lagrangian
          \item Coordinate descent steps in $z$ are available in closed form 
        \end{itemize}
      \item If $R^{(t)} : = \beta^{(t)}_{k^{\star}j^{\star}} - g(\beta^{(t)}_{k^{\star}})$ satisfies $|R^{(t)}| < \gamma$ or $t = t_{max}$, exit and return $\beta^{(t)}$
      \item Otherwise, set $\rho^{(t + 1)} = \tau\rho^{(t)}$ and $u^{(t + 1)} = \rho^{(t + 1)}R^{(t)}$
        \end{enumerate}
  \end{enumerate}
  \end{algorithm}
  
  \subsection{Constrained optimization via an efficient approximate Newton method}

As discussed in the main text (Section 3.2), we approximate the Hessian of the augmented Lagrangian induced by our constrained optimization problem as $-\nabla^2 l  + \rho (\nabla g - \vec{e}_{[k^{\star}j^{\star}]})(\nabla^T g - \vec{e}_{[k^{\star}j^{\star}]})$ and compute its inverse as a rank-one update to
$-(\nabla^2 l)^{-1}$ via the 
Sherman-Morrison identity.

We may justify this choice on several grounds. When $\rho$ is small in early iterations of our algorithm, $u$ will also be small, 
and so provided $g$ is not too large (which in practice it is not), $-\nabla^2 l$ will well approximate $\nabla^2\mathcal{L}$. In later iterations, we expect to have $g$ close to zero and $\rho \gg |u|$,
so $ (u + \rho g) \nabla^2g$ will in general be small relative to $ \rho \nabla g \nabla^T g$. Of course, these are somewhat loose justifications, and we may also observe that,
as the sum of a positive definite and a positive semi-definite matrix, $-\nabla^2 l  + \rho \nabla g \nabla^T g$ is positive definite and hence a Newton update using this approximation to 
the Hessian 
will always take us in a descent direction. 

\section{SI: Type 1 Error \& Power: Additional Results} \label{si:sims}

Tables \ref{tab:t1e} and \ref{tab:power} contain empirical type I error rate and power results for the proposed robust score tests and robust Wald tests on simulated data described in Section 6. Table \ref{tab:fd} contains a summary of false discoveries from the permutation analysis described in Section 7. 

\begin{table}[H]
\centering
\begin{tabular}{lllrrr}
  \hline
Test & Distribution & $J$ & $n = 10$ & $n = 50$ & $n = 250$ \\ 
  \hline
Robust score test & Poisson & 10 & 0.050 & 0.050 & 0.062 \\ 
Robust score test & Poisson & 50 & 0.036 & 0.038 & 0.050 \\ 
Robust score test & Poisson & 250 & 0.044 & 0.050 & 0.046 \\ 
Robust score test & ZINB & 10 & 0.032 & 0.026 & 0.028 \\ 
Robust score test & ZINB & 50 & 0.008 & 0.038 & 0.048 \\ 
Robust score test & ZINB & 250 & 0.034 & 0.066 & 0.042 \\ 
Robust Wald test & Poisson & 10 & 0.096 & 0.046 & 0.056 \\ 
Robust Wald test & Poisson & 50 & 0.146 & 0.050 & 0.056 \\ 
Robust Wald test & Poisson & 250 & 0.160 & 0.068 & 0.048 \\ 
Robust Wald test & ZINB & 10 & 0.242 & 0.058 & 0.030 \\ 
Robust Wald test & ZINB & 50 & 0.268 & 0.100 & 0.064 \\ 
Robust Wald test & ZINB & 250 & 0.316 & 0.112 & 0.058 \\ 
   \hline
\end{tabular}
\caption{Empirical type 1 error at the 5\% level for the robust score test and robust Wald test. Results are stratified by test, simulated distribution, number of taxa $J$, and sample size $n$. Results are shown for 500 simulations. }
\label{tab:t1e}
\end{table}  

\begin{table}[H]
\centering
\begin{tabular}{lcllrrr}
  \hline
Test & Signal magnitude & Distribution & $J$ & $n = 10$ & $n = 50$ & $n = 250$ \\ 
  \hline
Robust score test & 1 (small) & Poisson & 10 & 0.916 & 1.000 & 1.000 \\ 
Robust score test & 1 (small) & Poisson & 50 & 0.760 & 1.000 & 1.000 \\ 
Robust score test & 1 (small) & Poisson & 250 & 0.634 & 0.998 & 1.000 \\ 
Robust score test & 1 (small) & ZINB & 10 & 0.026 & 0.150 & 0.802 \\ 
Robust score test & 1 (small) & ZINB & 50 & 0.046 & 0.214 & 0.882 \\ 
Robust score test & 1 (small) & ZINB & 250 & 0.048 & 0.196 & 0.870 \\ 
Robust score test & 5 (large) & Poisson & 10 & 0.922 & 1.000 & 1.000 \\ 
Robust score test & 5 (large) & Poisson & 50 & 0.988 & 1.000 & 1.000 \\ 
Robust score test & 5 (large) & Poisson & 250 & 0.922 & 1.000 & 1.000 \\ 
Robust score test & 5 (large) & ZINB & 10 & 0.114 & 0.978 & 1.000 \\ 
Robust score test & 5 (large) & ZINB & 50 & 0.278 & 0.988 & 1.000 \\ 
Robust score test & 5 (large) & ZINB & 250 & 0.248 & 0.918 & 1.000 \\ 
Robust Wald test & 1 (small) & Poisson & 10 & 0.990 & 1.000 & 1.000 \\ 
Robust Wald test & 1 (small) & Poisson & 50 & 0.998 & 1.000 & 1.000\\ 
Robust Wald test & 1 (small) & Poisson & 250 & 0.988 & 1.000 & 1.000 \\ 
Robust Wald test & 1 (small) & ZINB & 10 & 0.304 & 0.282 & 0.818 \\ 
Robust Wald test & 1 (small) & ZINB & 50 & 0.330 & 0.408 & 0.912 \\ 
Robust Wald test & 1 (small) & ZINB & 250 & 0.394 & 0.432 & 0.902 \\ 
Robust Wald test & 5 (large) & Poisson & 10 & 1.000 & 1.000 & 1.000 \\ 
Robust Wald test & 5 (large) & Poisson & 50 & 1.000 & 1.000 & 1.000 \\ 
Robust Wald test & 5 (large) & Poisson & 250 & 1.000 & 1.000 & 1.000 \\ 
Robust Wald test & 5 (large) & ZINB & 10 & 0.926 & 1.000 & 1.000 \\ 
Robust Wald test & 5 (large) & ZINB & 50 & 0.938 & 1.000 & 1.000 \\ 
Robust Wald test & 5 (large) & ZINB & 250 & 0.964 & 1.000 & 1.000 \\ 
   \hline
\end{tabular}
\caption{Empirical power at the 5\% level for the robust score test and robust Wald test by test, signal magnitude, simulated distribution, number of taxa $J$, and sample size $n$. Results are shown for 500 simulations. The signal magnitude refers to the true value of $\beta_{2j^*}-g(\beta_2)$ when testing the null hypothesis $\beta_{2j^*}-g(\beta_2) = 0$. }
\label{tab:power}
\end{table}

\begin{table}[ht]
	\centering
	\begin{tabular}{|l|l|l|l|l|l|}
	  \hline
	Method & Minimum & First quartile & Median & Third quartile & Maximum \\ 
	  \hline
	ALDEx2 & $0$ & $0$ & $0$ & $0$ & $0$ \\
	ANCOM-BC2 & $299$ & $310$ & $316$ & $327$ & $340$ \\
  ANCOM-BC2 with SS & $6$ & $11$ & $14$ & $17$ & $28$ \\
	CLR t-test & $0$ & $0$ & $0$ & $0$ & $1$ \\
	DESeq2 & $0$ & $2$ & $4$ & $9$ & $14$ \\
	IFAA & $0$ & $0$ & $0$ & $0$ & $2$ \\
	radEmu score test & $0$ & $0$ & $0$ & $0$ & $0$ \\
	radEmu Wald test & $15$& $22$& $27$& $31$& $46$\\
	   \hline
	\end{tabular}
	\caption{\label{table:fdr} The number of false discoveries of each method determined via a permutation-based approach applied to the data of \cite{wirbel2019meta}, with 845 total taxa and a false discovery rate of 10\%. The data was constructed in order that all discoveries are false. Summaries of the distribution across 20 permutations are reported.}
	\label{tab:fd}
	\end{table}






\bibliographystyle{Chicago}
\bibliography{davids-papers-library} 

\end{document}